\documentstyle[11pt,psfig,leqno]{article}

\newcommand{\ud}{\mathrm{d}}
\newcommand{\half}{\frac{1}{2}}

\newcommand{\bem}{\begin{displaymath}}
\newcommand{\eem}{\end{displaymath}}

\newcommand{\beq}{\begin{equation}}
\newcommand{\eeq}{\end{equation}}
\newcommand{\bey}{\begin{eqnarray}}
\newcommand{\eey}{\end{eqnarray}}

\def\p{\partial}

\begin{document}

\title{A three-dimensional multidimensional gas-kinetic scheme for the Navier-Stokes equations under gravitational
fields}
\author{
C.L. Tian\thanks{National Astronomical Observatories, Chinese
Academy of Sciences, Beijing.  email: cltian@bao.ac.cn} ,
   K. Xu\thanks{Department of Mathematics,
      Hong Kong University of Science and Technology,  Hong Kong. email: makxu@ust.hk} ,
      K.L. Chan\thanks{Department of Mathematics,
      Hong Kong University of Science and Technology,  Hong Kong. email:
      maklchan@ust.hk},
L.C. Deng\thanks{National Astronomical Observatories, Chinese
Academy of Sciences, Beijing. email: licai@bao.ac.cn}     }

\date{}
\maketitle

\noindent{keywords: gas-kinetic scheme, Navier-Stokes equations,
gravitational field, compressible convection}

\section*{Abstract}

 This paper extends the gas-kinetic scheme for one-dimensional inviscid
 shallow water equations
 (J. Comput. Phys. 178 (2002), pp. 533-562) to  multidimensional
 gas dynamic equations under gravitational fields.
Four important issues in the construction of a well-balanced scheme
for gas dynamic equations are addressed. First,  the inclusion of
the gravitational source term into the flux function is necessary.
Second, to achieve second-order accuracy of a well-balanced scheme,
the Chapman-Enskog expansion of the Boltzmann equation with the
inclusion of the external force term is used. Third, to avoid
artificial heating in an isolated system under a gravitational
field, the source term treatment inside each cell has to be
evaluated consistently with the flux evaluation at the cell
interface. Fourth, the  multidimensional approach with the inclusion
of tangential gradients in two-dimensional and three-dimensional
cases becomes important in order to maintain  the accuracy of the
scheme. Many numerical examples are used to validate the above
issues, which include the comparison between the solutions from the
current scheme and the Strang splitting method. The methodology
developed in this paper can also be applied to other systems, such
as semi-conductor device simulations under electric fields.

\section{Introduction}

Since most astrophysical problems are related to the hydrodynamical
evolution in a  gravitational field, the correct implementation of
the gravitational force in an astrophysical hydrodynamical code is
essential, especially in systems with  long time integrations, such
as modeling star and galaxy  formation. Even though many
hydrodynamical codes have been successfully applied to astrophysical
problems, including the Piecewise parabolic method (PPM) and total
variation diminishing (TVD) codes \cite{frenk,ryu}, most have
considered only short time evolutions with strong shock or expansion
waves. With the slowness of galaxy evolution, many codes have
difficulties due to the improper treatment of the gravitational
force effect, the so-called source term in the Euler or the
Navier-Stokes equations. After integration of millions of time
steps, many codes cannot settle down to determine an isothermal
steady state solution for a gas in a time-independent external
gravitational field. The solution will either oscillate around the
equilibrium solution, or simply deviate from equilibrium due to
artificial heating, which triggers
 numerical gravitation-thermal instability, i.e., the collapse of
the gas core. There have been many attempts to construct such a
well-balanced gas dynamic code that preserves the hydrostatic
solution accurately \cite{leveque,zingale,botta}.

In recent years, the study of flow equations with source terms has
attracted much attention in the  computational fluid dynamics
community \cite{Toro}. Flow equations with source terms in
one-dimensional (1D) case can be written as
\begin{equation}
W_t + F(W)_x = S, \label{eq:1}
\end{equation}
where $W$ is the vector of conservative flow variables with the
corresponding fluxes $F(W)$. Here, $S$ is the source, such as the
gravitational force. If we integrate the above equation with respect
to $dx$ in a spatial element, $j$, from $x_{j-1/2}$ to $ x_{j+1/2}$,
and $dt$ in a time interval  from $t^n$ to $t^{n+1}$,
\begin{displaymath}
\int_{t^n}^{t^{n+1}} \int_{x_{j-1/2}}^{x_{j+1/2}} (W_t + F(W)_x ) dx
dt = \int_{t^n}^{t^{n+1}} \int_{x_{j-1/2}}^{x_{j+1/2}} S(W) dx dt,
\end{displaymath}
we get
\begin{displaymath}
W_j^{{n+1}} - W_j^{{n}} = \frac{1}{\Delta x} \int_{t^n}^{t^{n+1}}
\left( F_{j-1/2} (t) - F_{j+1/2} (t) \right) dt + \frac{1}{\Delta x}
\int_{x_{j-1/2}}^{x_{j+1/2}} \int_{t^n}^{t^{n+1}} S(W) dx dt ,
\end{displaymath}
where $\Delta x= x_{j+1/2} - x_{j-1/2}$ is the cell size and $W_j$
is the averaged value of $W$ in cell $j$, {\sl i.e.,}
$$ W_j = \frac{1}{\Delta x} \int_{x_{j-1/2}}^{x_{j+1/2}} W dx .$$
The above equation is an integral form of Eq.(\ref{eq:1}) and is
exact and equivalent to  Eq.(\ref{eq:1}).  In an earlier paper on
the shallow water equations \cite{xu3}, we emphasized the importance
of including the source term effect into the flux function at a cell
interface in order to develop a well-balanced scheme. However,  the
exact equilibrium solution, such as $u=0$ and $h+B=\mbox{constant}$,
for the shallow water equations, is relatively simple and special.
The simplicity can be realized through the relation $dh/dx=-dB/dx =
\mbox{constant}$ inside each computational cell once the bottom
topology is approximated as a linear function inside each cell.
Therefore, the MUSCL-type scheme with a single slope inside each
cell can be precisely used to reconstruct the initial well-balanced
solution. However, these techniques developed for the shallow water
equations cannot be directly extended to gas dynamics equations. For
example,
 the hydrostatic equilibrium solution for the gas dynamic equations
 may become $\rho \sim e^{-\alpha x}$, which is an exponential function.
 This exact equilibrium solution cannot even  be precisely reconstructed as
 an initial condition in  the MUSCL-type scheme with a single limited
 slope inside each numerical cell. Even though, it seems
 impossible to develop a well-balanced scheme for the gas dynamic
 equations with an error up to the order of
 machine zero, a proper numerical
 treatment is still essential to get accurate solutions.

 For a
gas flow under an external time-independent gravitational field,
there exists an isothermal solution, the so-called hydrostatic or
well-balanced equilibrium solution, with a constant temperature and
zero fluid velocity. This solution is an intrinsic solution due to
the balance between the flux gradient and source term, i.e., $F(W)_x
= S$. However, based on the macroscopic equations, it is unclear how
the gas settles down to an isothermal solution. Usually, additional
assumptions, such as a constant temperature, are needed before
deriving such an equilibrium result. Numerically, it is challenging
to obtain such a solution. If Eq.({\ref{eq:1}) is solved by  simple
operator splitting method, such as  $W_t + F_x =0$ for the flux at a
cell interface and $W_t = S$ to account for the source term inside
each cell, as observed frequently, many schemes simply deviate  from
the exact solution after millions of time step integrations. The
systematic errors will accumulate and eventually crash the
calculation. This happens to be the so-called \emph{numerical
heating} phenomenon, which was reported in \cite{slyz}. Due to the
accumulation of numerical errors, the internal energy in an isolated
gas system with adiabatic boundary conditions under a constant
gravitational field will increase forever. Figure~\ref{Figheat}
shows such a solution, where the internal energy goes up after
millions of time steps. The non-conservation of the total energy in
the system is not surprising because the gas dynamic equation
(\ref{eq:1}) itself is not written in a conservative form. The
problem in the gas dynamic equations is that the numerical error
accumulates in one direction. On the contrary, there is no such
problem in the shallow water equations even if the numerical scheme
is not a well-balanced one.

Based on  the Boltzmann equation, a flow  under a gravitational
field in a hydrostatic state can be described as
\begin{displaymath}
 {\vec c} \cdot \frac{\p f}{\p {\vec x}} -
\frac{\p \phi}{\p {\vec x}} \cdot \frac{\p f}{\p {\vec c}} = 0,
\end{displaymath}
where $f$ is the gas distribution function, ${\vec c}$ is the
particle velocity, and $\phi$ is the external gravitational
potential. The general solution of the above differential equation
is
$$ f({\vec x},{\vec c}) = {\cal F} (\phi + \half c^2 ) ,$$
where $\cal F$ is an arbitrary function of $c^2 = {\vec c}\cdot
{\vec c} = c_1^2 + c_2^2 + c_3^2 $. This equilibrium solution
clearly shows that the gas will have the same temperature, because
the temperature becomes a constant multiplier for the function $\phi
+ \half c^2$. For example, the function $\cal F$ in hydrostatic
equilibrium will become $\mbox{exp}(-(\half c^2 + \phi)/kT)$, where
$kT$ is a constant temperature. The above analysis shows the
usefulness of using a kinetic equation instead of macroscopic
equations to describe a gas flow under a gravitational field. It
also implicitly demonstrates that the flux functions, which are the
moments of $f$, have to take into account the source term effect,
i.e., $\phi$, in the solution of $f$.

In the past several years, a gas-kinetic BGK Navier-Stokes (BGK-NS)
scheme has been developed for compressible Navier-Stokes equations
\cite{xu2,li,may}.
 It achieves success in computing viscous flow solutions, such as
the hypersonic viscous and heat conducting flows
 \cite{sun}. Due to the simplicity of particle
transport, a multi-dimensional scheme can be constructed by
incorporating a multidimensional particle propagation mechanism
through the cell interface, where the particle transport in both
normal and tangential directions of a cell interface can be taken
into account.  However, constructing a well-balanced BGK-NS method
for gas dynamic equations under a gravitational field is not
straightforward. As shown above, with a hydrostatic state, the
density inside each numerical cell is distributed exponentially in
space, which cannot be precisely reconstructed  by a simple limited
slope in a standard MUSCL-type scheme. In other words, even starting
from the initial condition, some numerical error is already
introduced. Another important issue that needs to be addressed is
the Chapman-Enskog expansion of the Boltzmann equation, where the
external force effect has to be added to account for the balance
between the dissipative and external forcing terms. This is
important to capturing the viscous flow solution, especially with a
scheme that should work under both continuous and discontinuous flow
conditions. To achieve a well-balanced scheme, the gravitational
force effect on the particle acceleration and on the numerical flux
at the cell interface needs to be explicitly taken into account as
well. In order to avoid the numerical heating (cooling) effect, a
consistent treatment between the source term inside each cell and
the flux at the cell interface is required. The BGK scheme presented
in this paper is a well-balanced scheme up to second-order accuracy.
On the contrary to the shallow water solutions, the hydrostatic
equilibrium state for the gas dynamic equations, such as zero flow
velocity, cannot be precisely kept by the current scheme up to the
machine zero, i.e., $10^{-15}$. How to extend a well-balanced
strategy from the shallow water equations to the gas dynamic
equations remains an open problem.

This paper is organized as follows.  Section 2 gives the details of
the multi-dimensional BGK-NS scheme under a gravitational field.
Section 3 presents the results from many test cases, which confirm
the importance of some strategies in the development of a
well-balanced scheme. At the same time, the numerical heating
phenomenon is presented. The functions from the nonlinear limiter
and the multi-dimensional transport mechanism on the well-balanced
scheme are described. The last section is the conclusion.

\section{BGK scheme in three-dimensional space}

In this section, the multidimensional finite volume gas-kinetic
scheme based on the
 Bhatnagar-Groos-Krook (BGK) model for the compressible
 Navier-Stokes equations \cite{xu2} is extended to solve a flow system
 under an external gravitational force.
The governing equations solved by the gas-kinetic scheme are written
as,
\begin{equation}
\label{nses} \frac{\partial}{\partial t}\int_{\Omega}\vec{U}\ud
\Omega+\oint_{\vec{S}}\vec{F}\cdot\ud
\vec{S}=\int_{\Omega}\vec{Q}\ud \Omega,
\end{equation}
where $\Omega$ is the control volume, $\vec{S}$ is the surface of
this volume with its surface normal vector  pointing outward. The
macroscopic variables are
$$\vec{U}(\vec{x},t)=(\rho,\rho \vec{u}, E)^T, $$
where $\rho$ is the density, $\vec{u}$ is the velocity, and $E$ is
the sum of the thermal and kinetic energy. The fluxes take the form,
\begin{eqnarray*}
\vec{F}(\vec{x},t)=\left(\begin{array}{c}
\rho \vec{u}\\
\rho \vec{u}\otimes\vec{u}+p\bar{\bar{\mathrm{I}}}-\bar{\bar{\Sigma}}\\
(E+p)\vec{u}-\kappa\nabla T-\bar{\bar{\Sigma}}\cdot\vec{u}
\end{array}\right) ,
\end{eqnarray*}
where $\otimes$ denotes the tensor product of the vectors, $\kappa$
is the thermal conductivity coefficient, $\bar{\bar{\mathrm{I}}}$ is
the unit tensor, and $\bar{\bar{\Sigma}}$ stands for the viscous
shear stress tensor,
\begin{displaymath}
\Sigma_{ij}=\mu(\partial_i u_j+\partial_j u_i)-\varsigma(\nabla
\cdot\vec{u})\delta_{ij},
\end{displaymath}
where $\mu$ and $\varsigma$ are the first and second viscosity
coefficients. The source terms can be written as  follows:
\begin{displaymath}
\vec{Q}=(0,-\rho\nabla\phi,-\rho\vec{u}\cdot\nabla{\phi})^T,
\end{displaymath}
where $\phi=\phi(\vec{x},t)$ is the gravitational potential.

%%%

The BGK scheme is a finite volume method. In  Cartesian coordinates,
equation (\ref{nses}) is discretized as
\begin{equation}
\label{dscri}
\vec{U}_j^{n+1}=\vec{U}_j^{n}-\frac{1}{x_{j+1/2}-x_{j-1/2}}\int^{t^{n+1}}_{t^n}(\vec{F}_{j+1/2}(t)-\vec{F}_{j-1/2}(t))\ud
t+\int^{t^{n+1}}_{t^n}\vec{Q}_j\ud t,
\end{equation}
where $\vec{U}_j$ and  $\vec{Q}_j$ represent the cell-averaged
quantities  in the \emph{j}th cell, and $\vec{F}_{j+1/2}(t)$ is the
vector of  time-dependent fluxes across the cell interface between
the \emph{j}th and \emph{j+1}th cells.

 Generally, there are three stages in a shock capturing finite volume scheme, i.e.,
  reconstruction of initial data, gas evolution, and projection.
  In the reconstruction stage, a piecewise continuous
  flow distribution inside each cell is obtained. Then, in the gas evolution stage,
 a time-dependent flux, $\vec{F}_{j+1/2}(t)$, is computed based on the local solution at
 the
 cell interface, such as the solution of a Riemann problem.
 Based on  the fluxes, the  cell average
 flow quantities can be updated to a new time level. At the same
 time, the source term contribution has to be added on the time evolution of the flow
 variables inside each cell.

\subsection{Reconstruction}

In the higher order finite volume method, the cell-averaged flow
variables have to be reconstructed first to obtain the quantities at
the cell boundaries. The reconstruction depends solely on the flow
distribution at the beginning of each time step. In the smooth flow
region, a direct connection of cell averages across a cell interface
is a good choice,
\begin{equation}
\label{rc1} \vec{U}_{j+1/2}=\frac{\vec{U}_{j+1}+\vec{U}_{j}}{2},
\end{equation}
where $\vec{U}_{j+1/2}$ stands for the flow variables at the
interface, and the slope across the cell interface becomes
$({\vec{U}_{j+1}-\vec{U}_{j}})/(x_{j+1}-x_j)$.  A higher-order
interpolation can be used to improve the spatial accuracy of the
scheme. In a discontinuous flow region, the \emph{kinematic }
dissipation is  added implicitly to the reconstructed initial data
by converting kinetic energy into thermal one \cite{xu2}. The amount
of \emph{kinematic} dissipation depends on the limiters used. In the
current study, the van Leer limiter is adopted. For the sake of
simplicity, let $\vec{U}_{j+1}(x_{j+1/2})$ denote the reconstructed
data on the right-hand side of the interface ($x_{j+1/2}$) and let
$\vec{U}_{j}(x_{j+1/2})$  be the data on the left-hand side. Then,
\begin{equation}
\label{rc2} \vec{U}_{j}(x_{j+1/2})=\vec{U}_{j} (x_j) +
(x_{j+1/2}-x_j) L_j ,
\end{equation}
\begin{equation}
\label{rc3} \vec{U}_{j+1}(x_{j+1/2})=\vec{U}_{j+1} (x_{j+1})
-(x_{j+1}-x_{j+1/2}) L_{j+1}.
\end{equation}
In the above expressions, the van Leer limiter is defined as,
\begin{displaymath}
L_j=S(s_+,s_-)\frac{|s_+||s_-|}{|s_+|+|s_-|},
\end{displaymath}
where $S(s_+,s_-)=\mathrm{sign}(s_+)+\mathrm{sign}(s_-)$,
$s_+=(\vec{U}_{j+1}-\vec{U}_j)/(x_{j+1}-x_j)$, and
$s_-=(\vec{U}_{j}-\vec{U}_{j-1})/(x_{j}-x_{j-1})$.

\subsection{Gas evolution}

In the BGK scheme, the flow evolution  is governed by the BGK
equation \cite{bhatnagar},
 \begin{equation}
\label{bgk} \frac{\partial f}{\partial t}+\vec{c}\cdot\nabla
{f}+(-\nabla\phi)\cdot\nabla_{\vec{c}}f =\frac{g-f}{\tau},
\end{equation}
where $f(\vec{x},\vec{c},t)$ is the gas distribution function,
$\vec{c}$ is the particle velocity, $\tau$ is the collision time,
and $\nabla_{\vec{c}}=(\partial/\partial c_1,\partial/\partial
c_2,\partial/\partial c_3)$. The right-hand side of (\ref{bgk}) is
the so-called relaxation model, which is a simplification of the
collision term of the Botlzmann equation \cite{vincent}. The
equilibrium state, $g$, in equation (\ref{bgk}) is a Maxwellian
distribution,
\begin{displaymath}
g=\rho\left(\frac{\lambda}{\pi}\right)^{\frac{N+3}{2}}e^{-\lambda((\vec{c}-
\vec{u})\cdot(\vec{c}- \vec{u})+\xi^2)},
\end{displaymath} where $\xi$ has $N$
internal degrees of freedom, such as the molecular rotation,
$\lambda$ is related to the gas temperature by $\lambda=m/2kT$, $m$
is the molecular mass, $k$ is the Boltzmann constant, and $T$ is the
temperature. The above expression is for a gas in three dimensions.
For 1D and 2D flow simulations, the degree of freedom $K$ of the
variable $\xi$ is defined as $K=N+2$ for 1D flow and $K=N+1$ for 2D
flow, where $2$ stands for the particle random motion in the $x_2$
and $x_3$ directions and $1$ accounts for the $x_3$ direction random
motion only. In a 3D simulation, $K$ is equal to $N$. In the
equilibrium state, the internal energy variable, $\xi^2$, is defined
as $\xi^2=\xi^2_1+\xi^2_2+\ldots+\xi^2_K$.

The relation between the macroscopic variable, $\vec{U}$, and the
distribution function, $f$, is
\begin{eqnarray*}
\vec{U}=\int_{-\infty}^{+\infty}\vec{\psi} f\ud \Xi, \qquad
\vec{\psi}=[1,\vec{c},\frac{1}{2}(\vec{c}\cdot\vec{c}+\xi^2)]^T,
\end{eqnarray*}
where $\ud \Xi=\ud \vec{c}\ud \xi$ is the volume element in the
phase space with $\ud \xi=\ud\xi_1\ud\xi_2\ldots\ud\xi_K$. The
moments of a Maxwellian are given in Appendix A of \cite{xu2}.

 The Navier-Stokes equations
can be regarded as a low-order approximation of the BGK equation for
the linearly distributed flow variables in space with non-zero
velocity and temperature gradients. It has been shown that for a
monoatomic \cite{vincent} and polyatomic gas \cite{xu1}, the
Navier-Stokes equations can be derived from the BGK model with the
dissipative coefficients,
\begin{eqnarray*}
\label{ntc} \mu=\tau p,\qquad\varsigma
=\frac{2}{3}\frac{N}{N+3}\tau p,\qquad
\kappa=\frac{N+5}{2}\frac{k}{m}\tau p,
\end{eqnarray*}
which are  the dynamical viscosity coefficient, the second viscosity
coefficient, and the thermal conductivity, respectively. For an
ideal gas, the equation of state becomes $p=\rho/2\lambda$.

 The right-hand side
of equation (\ref{bgk}) describes  a relaxation process from an
initial non-equilibrium state, $f$, to an equilibrium state, $g$,
through particle collisions. Here, $\tau$  is  called the particle
relaxation time. Assume that $\tau$ is a local constant and
integrate equation (\ref{bgk}) along its particle trajectory,
\begin{displaymath}
\label{cl1} \frac{\ud \vec{x}'}{\ud t'}=\vec{c}'-\nabla{\phi}
t',\qquad \frac{\ud \vec{c}'}{\ud t'}=-\nabla{\phi}.
\end{displaymath}
We have
\begin{equation}
 \label{bgksln}
f(\vec{x},\vec{c},t)=\frac{1}{\tau}\int^t_0 ge^{-(t-t')/\tau}\ud t'
+e^{-t/\tau} f (\vec{x}_0,\vec{c}_0,0),
\end{equation}
where $0<t'<t$ and
\begin{displaymath}
\vec{c}_0=\vec{c}-(-\nabla\phi)t ,\qquad
\vec{x}_0=\vec{x}-\vec{c}t+\frac{1}{2}(-\nabla \phi)t^2
\end{displaymath}
are respectively the particle velocity and trajectory with initial
position $\vec{x}_0$ and velocity $\vec{c}_0$. The gravitational
effect has been included in the particle acceleration and trajectory
change. Since the term $\half (-\nabla \phi)t^2 $ has a third-order
effect on a numerical scheme, it will be ignored in the construction
of the time-dependent solution, $f$, in the following.

The integral of the right-hand side of equation (\ref{bgksln}) is
generally difficult to evaluate because the equilibrium state, $g$,
itself is implicitly related to $f$, and  the initial distribution,
$f(\vec{x}_0,\vec{c}_0,t'=0)$, is  unknown. In the BGK scheme, both
are modeled as first-order expansions of a local equilibrium
distribution with the inclusion of the Chapman-Enskog expansion for
the dissipative term. To include the effect of gravitational force
in the non-equilibrium gas distribution, the initial distribution,
$f_0=f(\vec{x}_0,\vec{c}_0,t'=0)$, is further developed as
\begin{eqnarray}
\label{inid}
 f_0 = \left\{
\begin{array}{l}
g_0^l[1-\vec{a}^l\cdot\vec{c}t-\vec{b}^l\cdot(-\nabla{\phi})t
-\tau(\vec{a}^l\cdot\vec{c}+\vec{b}^l\cdot(-\nabla{\phi})+A^l)],\quad x\le0\\
g_0^r[1-\vec{a}^r\cdot\vec{c}t-\vec{b}^r\cdot(-\nabla{\phi})t
-\tau(\vec{a}^r\cdot\vec{c}+\vec{b}^r\cdot(-\nabla{\phi})+A^r)],\quad
x\ge0,
\end{array}
\right.
\end{eqnarray}
where
\begin{displaymath}
g^l_0=\rho_{j}(x_{j+1/2})\left(\frac{\lambda_j(x_{j+1/2})}{\pi}\right)^{\frac{N+3}{2}}e^{\{
-\lambda_j(x_{j+1/2})((\vec{c}- \vec{u}_j(x_{j+1/2}))\cdot(\vec{c}-
\vec{u}_j(x_{j+1/2}))+\xi^2) \} }
\end{displaymath}
and
\begin{displaymath}
g^r_0=\rho_{j+1}(x_{j+1/2})\left(\frac{\lambda_{j+1}(x_{j+1/2})}{\pi}\right)^{\frac{N+3}{2}}e^{\{-\lambda_{j+1}(x_{j+1/2})((\vec{c}-
\vec{u}_{j+1}(x_{j+1/2}))\cdot(\vec{c}-
\vec{u}_{j+1}(x_{j+1/2}))+\xi^2)\}}
\end{displaymath}
 are the equilibrium distributions at the
left- and right-hand sides of the cell interface. The slopes in
equation (\ref{inid}) are defined as $\vec{a}^l=\nabla \ln g_0^l $,
$\vec{a}^r=\nabla \ln g_0^r $, $\vec{b}^l=\nabla_{\vec{c}} \ln
g_0^l$, $\vec{b}^r=\nabla_{\vec{c}} \ln g_0^r$, $A^l=\partial \ln
g_0^l/\partial t$, $A^r=\partial \ln g_0^r/\partial t$. The above
model is an extension of the Chapman-Enskog expansion in \cite{xu2},
where the gravitational force term has been added in the
modification of the non-equilibrium state. The term proportional to
$\tau$ corresponds to the dissipative terms in the Navier-Stokes
equations. The gravitational force makes two contributions on $f_0$.
One is the particle acceleration in the transport process to a cell
interface, and the other is in the non-equilibrium initial state.
Note that in the above initial gas distribution function,
 the flow variation in the tangential direction is also included in the definition of
 ${\vec a}^l $ and ${\vec a}^r$.

 All slopes in (\ref{inid}) can be uniquely evaluated from
 the reconstructed flow variables and their derivatives. In the normal
 direction of a cell
interface, ($x_{j+1/2}$), the slopes on the left-hand side  can be
expressed as
\begin{displaymath}
 a_j^l=\vec{\mathrm{a}}^l\cdot\vec{\psi},
\end{displaymath}
where $\vec{\mathrm{a}}^l$ is a local constant vector. Based on the
relation between the distribution function and the macroscopic
variables, we have
\begin{displaymath}
\frac{\vec{U}_{j}(x_{j+1/2})-\vec{U}_{j}(x_j)}{x_{j+1/2}-x_{j}}=\int
a_j^l\vec{\psi} g_0^l \ud \Xi =\vec{\mathrm{a}}^l\int
\vec{\psi}\otimes\vec{\psi}g_0^{l}\ud \Xi.
\end{displaymath}
The above linear system can be solved directly for $a_j^l$ and the
detailed formulation is given in the appendix for the 3D case.
 Similarly, the slopes on the right-hand side of the interface can be
obtained by solving
\begin{displaymath}
\frac{\vec{U}_{j+1}{(x_{j+1})-\vec{U}_{j+1}(x_{j+1/2})}}{
x_{j+1}-x_{j+1/2}}=\int a_j^r\vec{\psi}g_0^{r}\ud \Xi.
\end{displaymath}
In the tangential direction $(x_k)$ of the cell interface, the
slopes are obtained from
\begin{displaymath}
\frac{\vec{U}_{j, k+1}(x_{j+1/2})-\vec{U}_{j, k-1}(x_{j+1/2})}{
x_{j+1/2,k+1}- x_{j+1/2,k-1}}=\int a_k^l\vec{\psi}g_0^{l}\ud \Xi,
\end{displaymath}
\begin{displaymath}
\frac{\vec{U}_{j+1, k+1}(x_{j+1/2})-\vec{U}_{j+1, k-1}(x_{j+1/2})}{
x_{j+1/2,k+1}- x_{j+1/2,k-1}}=\int a_k^r\vec{\psi}g_0^{r}\ud \Xi.
\end{displaymath}

The terms multiplied by $\tau$ in equation (\ref{inid}) represent
the non-equilibrium part of a Chapman-Enskog expansion, and make no
direct contribution to the conservative flow variables. Therefore,
based on
\begin{eqnarray*}
\int(\vec{a}^l\cdot\vec{c}+\vec{b}^l\cdot(-\nabla{\phi})+A^l)\vec{\psi}
g_0^{l} \ud \Xi=0,\\
\int(\vec{a}^r\cdot\vec{c}+\vec{b}^r\cdot(-\nabla{\phi})+A^r)\vec{\psi}
g_0^{r} \ud \Xi=0,
\end{eqnarray*}
the temporal slopes $A^l$ and $A^r$ can be determined. Note that,
for the first time, the gravitational force effect on the
determination of  the non-equilibrium state has been proposed in the
above equations.

The equilibrium state $g(\vec{x}',\vec{c}',t')$ in the integral
solution  is constructed by a Taylor series expansion of a local
equilibrium distribution around $(\vec{x},\vec{c},0)$. With the
inclusion of the external source effects, it has the form
\begin{eqnarray}
\label{equ1} g= \left\{
\begin{array}{l}
g_0[1-(t-t')\bar{a}_n^lc_n-(t-t')\bar
{b}_n^l({-\nabla\phi})_n\\
\quad-(t-t')\bar{\vec{a}}_t\cdot\vec{c}_t-(t-t')\bar
{\vec{b}}_t\cdot({-\nabla\phi})_t+\bar{A}t'],\; c_n\le0\\
g_0[1-(t-t')\bar{a}_n^rc_n-(t-t')\bar
{b}_n^r({-\nabla\phi})_n\\
\quad-(t-t')\bar{\vec{a}}_t\cdot\vec{c}_t-(t-t')\bar
{\vec{b}}_t\cdot({-\nabla\phi})_t+\bar{A}t'],\; c_n\ge0
\end{array}
\right.
\end{eqnarray}
where
\begin{displaymath}
g_0=\rho_0\left(\frac{\lambda_0}{\pi}\right)^{\frac{N+3}{2}}e^{-\lambda_0((\vec{c}-
\vec{u}_0)\cdot(\vec{c}- \vec{u}_0)+\xi^2)}.
\end{displaymath}
The parameters in (\ref{equ1}) with subscript $n$ represent the
normal components of the vectors to the cell interface, and those
with subscript $t$ stand for the tangential components. This is
again an extension of $g$ in \cite{xu2}. It will be shown in Section
3 that the inclusion of both the external source term and the flow
gradients in the tangential direction in the expressions of $f_0$
(\ref{inid}) and $g$ (\ref{equ1}) is very important to obtain a
well-balanced scheme and to improve the accuracy and stability of
the scheme.

In order to determine all parameters in $g$ (\ref{equ1}), the
equilibrium state ($\vec{U}_0$) at the cell interface at $t=0$ must
be obtained first. Since the $f_0$ is totally determined, taking the
limit $t\rightarrow 0$ in equation (\ref{bgksln}) and substituting
its solution into conservation constraint $\int (g-f)\vec{\psi} \ud
\Xi=0$ at the cell interface, we have
\begin{eqnarray*}
\int g_0 \vec{\psi} \ud \Xi=\vec{U}_0=\int_{c_n>0}\int
g^l_0\vec{\psi}\ud \Xi+\int_{c_n<0}\int g^r_0\vec{\psi}\ud \Xi.
\end{eqnarray*}
Then, the slopes in equation (\ref{equ1}) can be obtained through
\begin{displaymath}
\frac{\vec{U}_0-\vec{U}_{j}(x_j)}{ x_{j+1/2}-x_{j}}=\int
\bar{a}_n^l\vec{\psi}g_0\ud \Xi,
\end{displaymath}
\begin{displaymath}
\frac{\vec{U}_{j+1}(x_{j+1})-\vec{U}_0}{x_{j+1}-x_{j+1/2}}=\int
\bar{a}_n^r\vec{\psi}g_0\ud \Xi,
\end{displaymath}
in the normal direction on the left- and right-hand sides of the
cell interface, and
\begin{displaymath}
\int \bar{a}_t\vec{\psi}\ud \Xi=\int_{c_n>0}\int
a_t^lg^l_0\vec{\psi}\ud \Xi+\int_{c_n<0}\int a_t^r
g^r_0\vec{\psi}\ud \Xi,
\end{displaymath}
in the tangential direction.

The only unknown in the expression of $g$ (\ref{equ1}) is $\bar{A}$.
By substituting the approximations (\ref{inid}) and (\ref{equ1})
into the general solution (\ref{bgksln}), we have
\begin{eqnarray}
f=&&\{(1-e^{-t/\tau})+(e^{-t/\tau}(t+\tau)-\tau)[(\bar{a}_n^lc_n+
\bar{b}_n^l({-\nabla\phi})_n)H[c_n]\nonumber\\
&&+(\bar{a}_n^rc_n+\bar{b}_n^r({-\nabla\phi})_n)(1-H[c_n])+\bar{\vec{a}}_t\cdot\vec{c}_t+
\bar{\vec{b}}_t\cdot({-\nabla\phi})_t]+[t-\tau(1-e^{-t/\tau})]
\bar{A}\}g_0\nonumber\\
&+&e^{-t/\tau}\{[1-\vec{a}^l\cdot\vec{c}t-\vec{b}^l\cdot({-\nabla\phi})t-\tau(\vec{a}^l\cdot\vec{c}+\vec{b}^l\cdot({-\nabla\phi})+A^l)]H[c_n]g_0^l\nonumber\\
&+&[1-\vec{a}^r\cdot\vec{c}t-\vec{b}^r\cdot({-\nabla\phi})t-\tau(\vec{a}^r\cdot\vec{c}+\vec{b}^r\cdot({-\nabla\phi})+A^r)](1-H[c_n])g_0^r\},
\end{eqnarray}
where the Heaviside function $H[x]$ is defined by
\begin{eqnarray*}
H[x]=\left\{
\begin{array}{l}
0,   x < 0, \\
1,   x\ge 0.
\end{array}
\right.
\end{eqnarray*}
Due to the conservation of the conservative flow variables during
particle  collisions, the moment of the right-hand side of equation
(\ref{bgk}) should vanish.  At the cell interface, the compatibility
condition gives
\begin{displaymath}
\int^{\Delta t}_{0}\int  (g-f)\vec{\psi} \ud t \ud \Xi=0,
\end{displaymath}
from which we get
\begin{eqnarray*}
\gamma_0\int \bar{A}\vec{\psi} g_0 \ud \Xi=&&\int[\gamma_1
g_0+\gamma_2 g_0((\bar{a}_n^lc_n+
\bar{b}_n^l({-\nabla\phi})_n)H[c_n]\nonumber\\
&&+(\bar{a}_n^rc_n+\bar{b}_n^r({-\nabla\phi})_n)(1-H[c_n])+\bar{\vec{a}}_t\cdot\vec{c}_t+
\bar{\vec{b}}_t\cdot({-\nabla\phi})_t)\nonumber\\
&+&\!\!\!\gamma_3(H[c_n]g_0^l+(1-H[c_n])g_0^r)\nonumber\\
&+&\!\!\!(\gamma_4+\gamma_5)(H[c_n](\vec{a}^l\cdot\vec{c}+\vec{b}^l\cdot({-\nabla\phi}))g_0^l\nonumber\\
&&+(1-H[c_n])(\vec{a}^r\cdot\vec{c}+\vec{b}^r\cdot({-\nabla\phi}))g_0^r)\nonumber\\
&+&\!\!\!\gamma_5(H[c_n]A^l g_0^l+(1-H[c_n])A^r g_0^r)]\vec{\psi}
\ud \Xi,
\end{eqnarray*}
where
\begin{eqnarray*}
\gamma_0&=&\int \tau(1-e^{-t/\tau})\ud t=\tau[\Delta t -\tau (1-e^{-\Delta t/\tau})],\nonumber\\
\gamma_1&=&\int e^{-t/\tau}\ud t=-\tau[e^{-\Delta t/\tau}-1],\nonumber\\
\gamma_2&=&\int -(e^{-t/\tau}(t+\tau)-\tau)\ud t=-\Delta t -(\gamma_4+\gamma_5), \nonumber\\
\gamma_3&=&\int -e^{-t/\tau}\ud t=-\gamma_1,\nonumber\\
\gamma_4&=&\int te^{-t/\tau}\ud t=-\tau[\Delta t e^{-\Delta t/\tau}-\tau(1-e^{-\Delta t/\tau})],\nonumber\\
\gamma_5&=&\int \tau e^{-t/\tau}\ud t=\tau \gamma_1.
\end{eqnarray*}
Therefore,  $\bar{A}$ is fully determined from the above equations.
As a result, the time-dependent gas distribution function,
$f(\vec{x},\vec{c},t)$, is obtained. The corresponding fluxes in the
$x_j$ direction across the cell interface can be computed by
\begin{eqnarray}
\label{flux1} \vec{F}_{j+1/2}&=&(F_{\rho},F_{\rho
\vec{u}},F_E)^T_{j+1/2}=\int c_j f\vec{\psi}\ud
\Xi\nonumber\\
 &=&\int[\Gamma_1
g_0+\Gamma_0\bar{A}g_0\nonumber\\
&+&\Gamma_2((\bar{a}_n^lc_n+
\bar{b}_n^l({-\nabla\phi})_n)H[c_n]\nonumber\\
&&+(\bar{a}_n^rc_n+\bar{b}_n^r({-\nabla\phi})_n)(1-H[c_n])+\bar{\vec{a}}_t\cdot\vec{c}_t+
\bar{\vec{b}}_t\cdot({-\nabla\phi})_t)g_0\nonumber\\
&+&\!\!\!\Gamma_3(H[c_n]g_0^l+(1-H[c_n])g_0^r)\nonumber\\
&+&\!\!\!(\Gamma_4+\Gamma_5)(H[c_n](\vec{a}^l\cdot\vec{c}+\vec{b}^l\cdot({-\nabla\phi}))g_0^l\nonumber\\
&&+(1-H[c_n])(\vec{a}^r\cdot\vec{c}+\vec{b}^r\cdot({-\nabla\phi}))g_0^r)\nonumber\\
&+&\!\!\!\Gamma_5(H[c_n]A^l g_0^l+(1-H[c_n])A^r g_0^r)]\vec{\psi}
c_j \ud \Xi,
\end{eqnarray}
where
\begin{eqnarray*}
\Gamma_0&=&\frac{1}{2}\Delta t^2 -\tau\Delta t+\tau (1-e^{-\Delta t/\tau}),\nonumber\\
\Gamma_1&=&\Delta t+\tau(e^{-\Delta t/\tau}-1),\nonumber\\
\Gamma_2&=&-\tau\gamma_2, \nonumber\\
\Gamma_3&=&-\tau\gamma_3,\nonumber\\
\Gamma_4&=&-\tau\gamma_4, \nonumber\\
\Gamma_5&=&-\tau\gamma_5.
\end{eqnarray*}

The above formulation is based on the discontinuous initial
condition at the cell interface, and the above scheme can be
faithfully applied to a flow with discontinuities. For a smooth
flow, based on the reconstruction without using the limiter, the
cell interface discontinuity will automatically disappear. So, under
this condition,   we have
\begin{displaymath}
\vec{a}^l=\vec{a}^r=\bar{\vec{a}},\quad
\vec{b}^l=\vec{b}^r=\bar{\vec{b}}, \quad A^l=A^r=\bar{A}.
\end{displaymath}
Therefore, the solution of the BGK equation (\ref{bgksln})  will
simply go to
\begin{eqnarray}
\label{fc}
f(\vec{x},\vec{c},t)&=&g(\vec{x},\vec{c},0)[1-\tau(\bar{\vec{a}}\cdot\vec{c}
+\bar{\vec{b}}\cdot(-\nabla \phi)+\bar{A})+\bar{A}t].
\end{eqnarray}
From this solution, we can clearly see that the BGK  solution from a
general discontinuous initial condition will go to the well-defined
continuous solution automatically, i.e., the Chapman-Enskog
expansion for a smooth flow under a gravitational field. In other
words, the multidimensional BGK scheme satisfies the consistency
requirement. Within a time scale, $\tau$, the particle acceleration
due to the gravity will distort the distribution from the
equilibrium through the term $\bar{\vec{b}}\cdot(-\nabla \phi)$.

It is well known that the BGK model has an intrinsic unit Prandtl
number. To make the BGK scheme simulate any heat-conducting flow,
the energy flux related to the heat conduction part in (\ref{flux1})
can be modified \cite{xu2} to
\begin{displaymath}
\label{newfe} \vec{F}_{E}^{new}=\vec{F}_{E}+(1/\mbox{Pr}-1){\vec q},
\end{displaymath}
where $\vec q$ is the heat flux in the normal direction of the cell
interface and $\mbox{Pr}$ is the Prandtl number. According to the
gas-kinetic theory, the heat flux is a result of energy transport
carried by molecules in their random movement.  In the $x_i$
direction, $\vec q$ can be evaluated by
\begin{eqnarray*}
q_i &=& \frac{1}{2} \int
(c_i-u_i)((\vec{c}-\vec{u})\cdot(\vec{c}-\vec{u})+\xi^2)f\ud \Xi.
\end{eqnarray*}
With a discontinuous initial condition,  the collision time has the
form \cite{xu2},
\begin{displaymath}
\tau=\frac{\mu_0}{p_0}+
\frac{|\rho_l/\lambda_l-\rho_r/\lambda_r|}{|\rho_l/\lambda_l+\rho_r/\lambda_r|}\Delta
t,
\end{displaymath}
where the first term comes from the Chapman-Enskog theory of the BGK
model, which states that the particle collision time is related to
the ratio of the dynamical viscosity coefficient to the pressure.
The second term enhances the dynamical dissipation in cases of flow
discontinuity. In the smooth flow region or around the slip line,
the numerical dissipation introduced above is very small because of
the continuous pressure distribution.

The above formulation clearly shows the complexity of the
gas-kinetic scheme for the Navier-Stokes solutions under a
gravitational field. The splitting of a Maxwellian distribution into
two half ones leads to a very complicated flux splitting method with
error functions and terms involving exponentials. The drawback may
be overcome by Perthame's compactly supported equilibrium
distribution function \cite{perthame} instead of a full Maxwellian
distribution. This will makes BGK scheme much simpler and
attractive, which deserve further investigation.

\subsection{Projection}

In the projection stage, the average mass, momentum, and energy
inside each cell are updated. Any inconsistent numerical treatment
between the flux and the source term in this step may easily cause
\emph{numerical heating} in the simulation of gravitational gas
dynamics.
 Here, we are going to analyze the so-called \emph{numerical
heating}  reported by Slyz \cite{slyz}. When  \emph{numerical
heating} occurs, the computed system will continuously increase its
internal energy, which is shown in Fig.~\ref{Figheat}.

Typically, there are two ways to update the energy inside each cell.
One is the  \emph{non-conservative} form:
\begin{eqnarray}
\label{ude1} \frac{\partial}{\partial t}\int_{\Omega}E\ud
\Omega+\oint_{\vec{S}}((E+p)\vec{u}-\kappa\nabla
T-\bar{\bar{\Sigma}}\cdot\vec{u})\cdot\ud
\vec{S}\nonumber\\
=\int_{\Omega}\rho\vec{u}\cdot(-\nabla\phi)\ud \Omega.
\end{eqnarray}
For a time-independent potential, the gravitational energy can be
absorbed into the total energy. Therefore,  the energy equation in
this special case can be written in a conservative form as the
following:
\begin{equation}
\label{ude2} \frac{\partial}{\partial t}\int_{\Omega}E_{tot}\ud
\Omega+\oint_{\vec{S}}((E_{tot}+p)\vec{u}-\kappa\nabla
T-\bar{\bar{\Sigma}}\cdot\vec{u})\cdot\ud \vec{S}=0,
\end{equation}
where $ E_{tot}= E +\rho\phi $. The discretized forms of equations
(\ref{ude1}) and (\ref{ude2}) are
\begin{equation}
\label{de1}
E_j^{n+1}=E_j^{n}-\frac{1}{x_{j+1/2}-x_{j-1/2}}\int^{t^{n+1}}_{t^n}(F_{E}(x_{j+1/2},t)-F_{E}(x_{j-1/2}
,t))\ud t+\int^{t^{n+1}}_{t^n}Q_E(x_j)\ud t,
\end{equation}
\begin{equation}
\label{de2}
E_{tot}^{n+1}(x_j)=E_{tot}^n(x_j)-\frac{1}{x_{j+1/2}-x_{j-1/2}}\int^{t^{n+1}}_{t^n}(F_{E_{tot}}(x_{j+1/2},t)-F_{E_{tot}}(x_{j-1/2},t))\ud
t,
\end{equation}
respectively, where the gravitational source term is
\begin{displaymath}
\int Q_E(x_j)\ud
t=\frac{1}{\Omega}\int\int_{\Omega}\rho\vec{u}\cdot(-\nabla\phi)\ud
\Omega\ud t.
\end{displaymath}

 Slyz and Prendergast concluded
that equation (\ref{de2}) rather than (\ref{de1}) should be used to
update the energy term. Otherwise,  \emph{numerical heating} will be
introduced into the numerical solution because of the inaccurate
computation of the source term \cite{slyz}. Eventually, the
numerical gravitational system will break down. This argument is
perfectly correct. However, it is not totally complete. With a
careful discretization of the source term in (\ref{de1}), we can
still establish a stable and accurate solution. Numerically, the
inaccuracy comes mainly from the low-order approximation and the
\emph{inconsistent} computation of the source term. Here,
\emph{inconsistency} means that the same flow variable at different
locations of the governing equations are calculated by different
methods. For example, in the projection formulation, if the gravity
 is time-independent, then the source term in equation (\ref{de1}) can
be approximated by
\begin{equation}
\label{nh1}
\frac{1}{\Omega}\int\int_{\Omega}\rho\vec{u}\cdot(-\nabla\phi)\ud
\Omega\ud t=\frac{1}{\Omega}\int\left[\int \rho \vec{u} \ud
t\right]\cdot(-\nabla\phi)\ud \Omega,
\end{equation}
where the term in the square brackets is the time integral of the
mass flux, which should be the same as the corresponding term in the
continuum equation.  In Slyz and Prendergast's paper \cite{slyz},
the mass flux in the continuous equation was calculated by an early
BGK scheme (denoted  as $\vec{F}_{\rho}^{bgk}$), while the mass flux
in equation (\ref{nh1}) was computed by a bilinear interpolation
(denoted  as $\vec{F}_{\rho}^{bi}$). Suppose their difference is
$\vec{F}_{\rho}^{bi}-\vec{F}_{\rho}^{bgk}=\delta \vec{Q}$. If all
the fluxes in (\ref{dscri}) are calculated by the BGK method while
the source term in (\ref{de1}) is calculated by
$\vec{F}_{\rho}^{bi}\cdot\frac{1}{\Omega}\int (-\nabla\phi)\ud
\Omega$, the additional source term, $\delta \vec{Q}\cdot
\frac{1}{\Omega}\int(-\nabla\phi)\ud \Omega$, will appear in the
energy equation,  (\ref{de1}). This could be the origin of
\emph{numerical heating}. Unfortunately, the error will grow in one
direction in a system under gravity. The instability is closely
related to the so-called {\sl gravo-thermal instability}. But the
reason is due to the numerical conservation error, instead of the
physical one due to the energy loss in an astrophysical system.

For an isolated or adiabatic system, after a long time integration,
any inconsistent treatment of the source term in the formulation
(\ref{de1}) will generate \emph{numerical heating} and the system
will deviate from its physical solution. In order to avoid the
instability and inaccuracy in the solution, we have taken the
following two steps when using (\ref{de1}). The first one is to
compute the source term consistently. In the current study, the mass
flux at the cell interface is used to approximate the integration
inside the cell. The  source term, (\ref{nh1}), is discretized as
\begin{displaymath}
\int Q_E(x_j)\ud t=\frac{1}{2}\int(
F^{bgk}_\rho(x_{j-1/2})+F^{bgk}_\rho(x_{j+1/2}))(-\nabla\phi)_j \ud
t.
\end{displaymath}
The second one is that the gravitational forcing effect is
explicitly included in the flux evaluation as presented for the BGK
scheme in the last subsection. Equipped with these two mechanisms,
the scheme with (\ref{de1}) is still a well-balanced method up to
the second-order accuracy.

When equation (\ref{de2}) is used, the time-integrated total energy
flux can be written as
\begin{displaymath}
\vec{F}_{E_{tot}}=\vec{F}_{E}+\vec{F}_{\rho}\phi.
\end{displaymath}
Therefore, the total energy will be precisely conserved. Even
through the gravitational force effect is not included in the flux
evaluation in \cite{slyz}, based on  the formulation (\ref{de2}) the
numerical error will not grow and accumulate forever. The departure
from the isothermal hydrostatic state will stay bounded due to the
total energy conservation, but with large oscillations.  Even when
using (\ref{de2}),
 the numerical accuracy
can be much improved once the gravitational force is included in the
flux construction. The numerical results presented in the next
section support the above argument.

\section{Numerical tests and discussion}

In this section, numerical tests of 1D, 2D and 3D cases are
performed to validate the gas-kinetic method. Each of them is very
sensitive to the accuracy of the scheme. Some of them are run for
millions of numerical steps whereby the accumulation of any small
error would become significant for such a long time integration.

\subsection{Perturbation of the 1D isothermal equilibrium solution}

This first test case is from the the paper by LeVeque and Bale
\cite{leveque}, where a quasi-steady wave-propagation algorithm was
developed for an ideal gas subject with a static gravitational
field. Initially, an ideal gas with $\gamma =1.4$ stays at an
isothermal hydrostatic state,
\begin{eqnarray*}
\rho_0(x)=p_0(x)=e^{-x}, \quad \mbox{and} \quad  u_0(x)=0,
\end{eqnarray*}
for $x\in [0,1]$. The gravity acts in the negative $x$ direction,
such as $ -\nabla\phi(x)=-1$. Initially, the pressure is perturbed
by
\begin{displaymath}
p(x,t=0)=p_0(x)+\eta e^{-\alpha(x-x_0)^2},
\end{displaymath}
where $\alpha=100, x_0=0.5$, and $\eta$ is the amplitude of the
perturbation.

The computation is conducted with $100$ grid points in the whole
domain. The results from the BGK scheme with and without the
gravitational source term included in the fluxes are shown in
Figure~\ref{FigLe1}. The figure clearly shows that the inclusion of
the source term in the flux function is very important for the
accurate capturing of small perturbations. In the above
calculations, the BGK scheme uses the van Leer limiter in the
initial reconstruction. Therefore, the BGK scheme itself works in
both the smooth and discontinuous flow regions. However, the
quasi-steady wave propagation algorithm in \cite{leveque} works only
in the smooth flow region.

\subsection{One-dimensional gas falling into a fixed external potential}

This case is taken from Slyz and Prendergast's paper \cite{slyz}  to
investigate the numerical accuracy of the BGK scheme. The gas is
initially stationary  ($\vec{u}=0$) and homogenous ($\rho=1,\;
\epsilon=1$, where $\epsilon$ is the internal energy). The
gravitational potential has the form of a sine wave,
\begin{displaymath}
\phi=-\phi_0\frac{\mathcal{L}}{2\pi}\sin{\frac{2\pi
x}{\mathcal{L}}},
\end{displaymath}
where ${\cal L}=64$ is the length of the computational domain and
$\phi_0=0.02$. The ratio of the specific heat, $\gamma$, is $5/3$.
 The
periodic boundary conditions are implemented in this system. After
the initial transition, the system eventually reaches an isothermal
hydrostatic distribution, where the temperature settles to a
constant ($T(x,t)=T_0$) and $\vec{u}=0$. For an ideal gas ($p=\rho
RT$),
 from the hydrostatic equation $\partial p/\partial x=\rho(-\partial \phi/\partial x) $,
  the density distribution becomes
   $\rho(x,t)\sim\exp{\left(-{\phi}/{RT_0}\right)}$. However, the system is
  highly nonlinear  since  $T_0$ depends on $\rho$. It is thus
better to use a numerical method
  to solve this problem.   Each simulation  result presented here is
  obtained with $500,000$ time steps on a Cartesian grid  with $\Delta x=1$.
  The Courant number used is $0.9$.

 When the \emph{numerical heating } occurs in an
inconsistent scheme, the gas will  become hotter due to the
numerical error.
 As used by Slyz, the schemes solving
equation (\ref{ude1}) are named the energy source term (EST)
schemes, and others solving equation (\ref{ude2}) are called the
energy conservation (ECT) schemes.  The final state (density,
velocity and temperature) of the system given in Figure~\ref{FigUe1}
is calculated from both the EST and ECT BGK schemes. A
\emph{consistent} source term treatment for the EST scheme is used,
such that the mass flux integration inside each cell is approximated
by the average of the cell interface mass fluxes. In the above
calculation,  the gravitational effect is included in the flux
evaluations for both schemes. In this figure, the diamonds indicate
 the solutions of the ECT scheme and the solid lines indicate the
solutions of the EST scheme. Figure~\ref{FigUe1} shows that the
results obtained by the EST scheme and ECT scheme are almost
indistinguishable. The small derivation from the hydrostatic state
in both schemes is due to the inaccurate representation of the
exponential function by a single slope inside each cell. We cannot
find any simple way to remedy this.

Even within the framework of the ECT scheme, in the following we are
going to show that it is necessary to  include the gravitational
force in the flux for maintaining high accuracy. For a stationary
state, $\vec{u}=0$, it can be shown that in the current scheme the
fluxes induced by the spatial derivatives
$\int\vec{a}\cdot\vec{c}c_ig\vec{\psi} f\ud \Xi$ are totally
balanced by the gravitational force terms
$\int\vec{b}\cdot(-\nabla\phi)c_ig\vec{\psi} f\ud \Xi$, where there
is no adjustable parameter.  In other words, without considering the
external force effect in the flux, the steady state cannot be
accurately reached by a numerical scheme. The solution given in
Figure~\ref{FigUe2} is calculated by ECT scheme. The diamonds are
the solution of the ECT scheme with a gravitational effect in the
flux and the solid lines are the ones  without the gravitational
effect in the flux. Figure~\ref{FigUe2} shows that neglect of the
gravitational effect in the gas evolution stage will drive the
solution away from the hydrostatic equilibrium solution even though
the solution will not blow up due to the total energy conservation,
the so-called ECT scheme.  We believe that any scheme based on the
Riemann solution, if the gravitational term is not included in the
flow evolution, may have a similar problem.  By including gravity,
the relative fluctuation in both the velocity and temperature is
much reduced. It is difficult to predict if there will be any shock
capturing well-balanced scheme, where the fluid velocity in a
hydrostatic equilibrium state can be kept up to the machine zero.

This test case illustrates that a simple operator splitting approach
does not work and the inclusion of the gravitational force in the
numerical flux is the key for a well balanced scheme. However, it is
widely recognized and proved mathematically that the Strang
splitting method is a second order accurate time-integration method
for general evolutionary equations with multiple operators. In the
following, we apply the Strang splitting method to this test case as
well. The update of the flow variables are
\begin{equation}
\vec{U}^{n+1}=S[\Delta t/2]H_{BGK}[\Delta
t]S[\Delta t/2]\vec{U}^{n},
\end{equation}
where $S$ represents the source term contribution inside each cell,
and $H_{BGK}$ is the BGK flux across a cell interface without
including the gravitational force term. The results from the EST BGK
scheme and the above EST Strang splitting method are shown in
Figure~\ref{FigStrang2}. The Strang splitting method gives accurate
velocity solution, which is similar to that from EST BGK.  However,
the temperature of the EST Strang solution is continuously
increasing with the total number of integration steps. For a slow
gas evolution system under gravitational field, the numerical
heating from the Strang splitting method will become severe.

\subsection{Onset of two-dimensional compressible convection}

Consider a fluid confined by two horizontal parallel plates, where
the gravity is imposed vertically, and the upper plate is fixed at a
lower temperature while the lower plate is fixed at a higher
temperature. For inviscid fluids, when the temperature gradient
exceeds the adiabatic temperature gradient, the system becomes
convectively unstable. In the astrophysics literature, such a
condition is known as the Schwarzschild criterion for convection
(\cite{stellar},p39). For general cases where the viscosity is taken
into account, the motion is always impeded by the viscosity, and the
temperature gradient is reduced by the thermal conductivity. To
study the general situation, the Rayleigh number is used to indicate
the combined effects of the dynamical viscosity, thermal diffusion
and temperature gradient. The steady convection will not occur
unless the Rayleigh number is larger than a specific value, i.e.,
the critical Rayleigh number.  The compressible convection is a
sensitive test on the accuracy and stability of any numerical
scheme. The onset of stable large-scale circulation depends
sensitively on the numerical dissipation and boundary treatment. A
low-order scheme may introduce too much artificial dissipation,
which increases the critical Rayleigh number. Due to the large scale
in height and pressure changes, the instability can be triggered
easily.

In the gas-kinetic BGK scheme, the dissipative coefficient  is
controlled  by the particle collision time. In a continuous flow
region, only physical viscosity needs to be used. However, there are
two ways to add the artificial dissipation in a numerical scheme to
enhance the shock-capturing ability. The \emph{kinematic}
dissipation can be introduced through limited reconstructed initial
data, which implicitly transfers kinetic energy into thermal energy.
The \emph{dynamical} dissipation can be added as well by enlarging
the viscosity coefficient in the governing equations according to
the pressure jump at the cell interface \cite{xu2}. Note that any
discontinuity in the reconstructed data means the cell resolution is
not enough to resolve the flow structure.
 Here, the evaluation of the critical Rayleigh numbers for the
 two-dimensional compressible convection case  is a good indication of
 the accuracy and
robustness of the scheme.

In this section, we adopt the cases calculated by Grahma
\cite{graham} and Chan \cite{chan1} for the purpose of comparison.
Consider an ideal gas confined in a box with $0\le x_1\le l,\;
0\le x_2\le d$. The free surface and isothermal boundary
conditions are imposed at the top and bottom boundaries,
\begin{displaymath}
u_2=0,\quad\frac{\ud u_1}{\ud x_2}=0,\quad T=T_b,T_t,\quad
\textrm{at}\quad x_2=0,d,
\end{displaymath}
and the stress-free insulated boundary conditions are imposed in the
horizontal direction,
\begin{displaymath}
u_1=0,\quad\frac{\ud u_2}{\ud x_1}=0,\quad \frac{\ud T}{\ud
x_1}=0,\quad \textrm{at}\quad x_1=0,l.
\end{displaymath}
The initial condition is the static state with a small perturbation
in the  velocity field,
\begin{eqnarray*}
T&=&(1+Z(d-x_2)/d)T_t,\\
\rho&=&(T/T_t)^m\rho_t,\\
p&=&(T/T_t)^{m+1}p_t,
\end{eqnarray*}
where $Z=(T_b-T_t)/T_t$ is the normalized parameter, and $m$ is the
polytropic gas index. There is a hydrostatic solution of the
Navier-Stokes equations with a  uniform constant gravitational
acceleration,
\begin{displaymath}
-\nabla\phi=-\frac{(m+1)ZRT_t}{d}.
\end{displaymath}
With the above $T, \rho$, and $p$ distributions,  the initial
vertical velocity is perturbed by
\begin{eqnarray*}
u_1&=&0,\\
u_2&=&u_0\sin{\frac{2\pi x_1}{l}}\sin{\frac{\pi x_2}{d}},
\end{eqnarray*}
where $u_0$ is a very small constant. Actually, the initial
perturbation for the vertical velocity may be randomly generated
with the satisfaction of the boundary condition. The above form
leads to a faster startup of the convective rolls. All the
quantities, $\rho_t$, $p_t$, $T_t$, $d$ and $l$, are normalized. In
all computations presented here, $\gamma=5/3 $, $Pr=1$ and $m=1.4$
are used. The Rayleigh number is defined as
\begin{eqnarray}
\label{rn}
 Ra=\frac{PrRT_tZ^2d^2\rho_t^2}{\mu^2}\left[\frac{1-(\gamma-1)m}{\gamma}\right](m+1).
 \end{eqnarray}
The standard solutions of the two dimensional laminar convection are
shown in Figure~\ref{Figsl}. A discussion of these solutions can be
found in \cite{chand}.

 For each $Z$, the calculation is repeated with two slightly
supercritical Rayleigh numbers, $Ra_1$ and $ Ra_2$. Then, the
critical Rayleigh number is determined by fitting the curve
$\max{(u_2)}=A\exp{(B(Ra-Ra_c)t)}$ by a least square approximation,
where $u_2$ is the vertical velocity, $Ra_c$ is the critical
Rayleigh number, and $A$ and $B$ are the constants to be determined.
The results  given in Table~\ref{crn} are obtained from the current
BGK scheme, where a continuous interpolation (\ref{rc1}) is used in
the reconstruction stage. From Table~\ref{crn}, we can see that the
current BGK scheme is an accurate method for a viscous flow under a
gravitational field. The difference between the critical Rayleigh
numbers calculated from the gas-kinetic BGK scheme and those from
the linear analysis is less than 3.5\% for $Z\le 10$ and $\leq 1\%$
for $Z<4$. Also, the Courant number used for each calculation is
included in Table \ref{crn}. When a nonlinear limiter,
Eq.(\ref{rc2}) and Eq.(\ref{rc3}), is used in the  reconstruction of
the initial data, discontinuities at the cell interfaces will be
introduced. This will generate extra {\sl kinematic} dissipation.
Certainly, the interface discontinuity will become smaller for
higher-order interpolation in the smooth region, as will the
numerical dissipation. But, instead of using higher-order
interpolation, the van Leer limiter is always used in this paper
because of its robustness and simplicity. To investigate the effect
of the limiter on the Rayleigh number, the gas-kinetic BGK scheme
with the van Leer limiter is used for the same calculation. The
numerical computation shows that, for the weakly compressible flow,
the critical Rayleigh number becomes slightly larger ($Ra=553.67$
for $Z=0.5$), while for highly compressible flow, the situation is
complex. The critical Rayleigh number becomes a little bit smaller
($Ra= 289.51$ ) for $Z=2$ and larger ($Ra= 167.5$) for $Z=4$. For
$Z=10$, the instability appears near the upper surface. This can be
explained by the use of a limiter, which continuously perturbs the
boundary layer through the introduction of numerical
discontinuities.  The limiter generates a jump in the vertical
velocity field.  A direct remedy for this is to use the
high-resolution interpolation, or modify the initial dissipative
terms in $f_0$ to make these terms continuous across the cell
interface \cite{may}. However, this error due to the limiter could
be ignored for small and  modest $Z$. In conclusion, a nonlinear
limiter will introduce
 numerical dissipation, but its effect on the calculation of the
 critical Rayleigh number is tolerable.
  Furthermore, for $Pr\ll1$ cases, the effect from
 the nonlinear limiter becomes negligible, where the heat conduction dominates
 and smoothes the flow field.

\subsection{Thermodynamic properties of the laminar convection in 3D}

In the numerical flux,  the derivatives of the flow distribution in
the tangential direction of a
 cell interface are taken into account in the current scheme.
 The current scheme is essentially a
multidimensional method, where the physical effect from both
gradients in the normal and tangential directions is taken into
account. In order to verify the necessity of this kind of
discretization, tests of compressible convection in a
three-dimensional box will be presented in this subsection.

The two-dimensional model in the previous subsection is extended to
three dimensions. The aspect ratio of the rectangular box is
1:0.1:1. The insulated solid well boundary conditions are applied in
the horizontal directions ($x_1$ and $x_2$). Initially, the velocity
field is perturbed in all directions: $x_1$, $x_2$ and $x_3$. The
Courant number takes a value of $0.3$ and the grid size is $50\times
5\times 50$. Unless otherwise specified, all the parameters in this
subsection have the same value as those in the previous subsection.

Among the solutions, the maximum Mach number and Nusselt number are
used to measure the thermodynamical property and convective heat
transfer. The maximum Mach number is defined as
\begin{displaymath}
Mc=\max{\left\{\frac{|u|}{\sqrt{\gamma R T}}\right\}},
\end{displaymath}
which sensitively depends on  the viscosity.

The Nusselt number is used to compare the flux carried by the heat
conduction and by convection. Following  Graham \cite{graham}, it is
defined as,
\begin{displaymath}
{\cal N} =\frac{F_t-F_a}{F_c-F_a},
\end{displaymath}
where $F_a=-\Delta\phi k/c_p$, $F_c=\kappa(T_b-T_t)/d$ are the heat
flux due to the adiabatic gradient and conduction, respectively. Let
\begin{displaymath}
F_t=\overline{\rho c_p T u_3+\kappa\frac{\ud T}{\ud x_3}}
\end{displaymath}
be the total heat flux, where the over-bar means the average over
the whole computational domain. Numerical experiments show that
${\cal N}$ cannot be correctly obtained with a low order scheme.
 We also introduce the relative Rayleigh number:
\begin{displaymath}
R^{\ast}=\frac{R}{Rc},
\end{displaymath}
where $R$ is defined by equation (\ref{rn}) and $Rc$ is the critical
Rayleigh number from the linear analysis.

A series of finite-amplitude steady solutions is calculated with
different Rayleigh numbers. The depth parameter is fixed to a unit.
The results of one-cell circulation are given in  Table~\ref{mc-rn}
and  Table~\ref{nu-rn}. The comparison with Graham's results is
shown in Figure~\ref{Figramn}.

From plot (a) in Figure~\ref{Figramn} we can see that the agreement
is good when the central interpolation is used for the initial
reconstruction (the average deviation $\bar{\sigma} \leq  9\%$). The
agreement is better for larger Rayleigh numbers, for example,
$\sigma(R^\ast=100) \leq 5\%$. This is expected since a BGK scheme
with a central interpolation is similar to the Lax-Wendroff scheme,
which was used by Graham \cite{graham}. As pointed out in the
previous section, the numerical dissipation will be introduced when
the van Leer limiter is adopted. The average deviation of the
results with different reconstructions is less than $2\%$. This kind
of discrepancy increases with the Rayleigh number
($\sigma(R^\ast=3)=1.4\%)$ and $\sigma(R^\ast=100)=3.4\%$).

The trend of the discrepancy in plot (b) of Figure~\ref{Figramn} is
similar to that in plot (a), except that the best fitting results
come with the van Leer limiter.  The average deviations from
Graham's results are approximately less than $3\%$ and $5\%$ using
the van Leer limiter and the central interpolation, respectively.
The Nusselt numbers obtained from the BGK scheme are systematically
larger than Graham's results.

The effects of including  tangential derivatives, i.e., the
so-called multi-dimensionality, into the numerical fluxes have been
 checked. The results obtained from a dimensional-splitting BGK
method,
\begin{displaymath}
\vec{U}^{n+1}_\ast=H_{x_1}\vec{U}^{n},
\end{displaymath}
\begin{displaymath}
\vec{U}^{n+1}_{\ast\ast}=H_{x_2}\vec{U}^{n+1}_\ast,
\end{displaymath}
\begin{displaymath}
\vec{U}^{n+1}=H_{x_3}\vec{U}^{n+1}_{\ast\ast}+\vec{Q}(\vec{U}^{n+1}_{\ast\ast}),
\end{displaymath}
are also listed in Tables~\ref{mc-rn} and \ref{nu-rn}, where
$H_{x_i}$ is the one-dimensional operator and $\vec{Q}$ is the
source term. In the computation of the numerical fluxes,  the
tangential derivatives are excluded. In the directional splitting
method, it is clear that additional numerical dissipation is added
to the scheme because the maximum Mach numbers in the third row of
Table~\ref{mc-rn} are systematically less than those in the second
row. The situation is complicated for the Nusselt number. When a
dimensional splitting method is used, the convection becomes
slightly turbulent for $R^\ast \geq 70$. The numbers with a
superscript $\ast$ represent temporally averaged values. The
standard deviation for $Mc(R^\ast=100)$, $N(R^\ast=70)$ and
$N(R^\ast=100)$ are 0.001, 0.003 and 0.226, respectively. When the
flow becomes turbulent, the Nusselt numbers from the
multidimensional scheme are a bit smaller than those from the
dimensional splitting scheme which indicates that the turbulence has
a more efficient way to transfer the heat than the laminar flow.
This test clearly shows that the multidimensional scheme is more
stable and accurate than the dimensional splitting scheme. The
reason lies in the fact that to keep both normal and tangential
derivatives in the evaluation of a flux function in a circulation
motion is close to the physical reality.

\subsection{Rayleigh-Taylor instability}

This test case is similar to LeVeque and Bale's Rayleigh-Taylor
instability case \cite{leveque}, which was used to validate their
quasi-steady wave propagation scheme.  Consider an isothermal
equilibrium ideal gas ($\gamma=1.4$) in a 2D polar coordinate,
\begin{eqnarray*}
\rho_0 (r)=e^{-\alpha(r+r_0)},\\
p_0(r)=\frac{1.5}{\alpha}e^{-\alpha(r+r_0)},\\
-\nabla \phi (r)=-1.5,\\
u_0=0,\\
\end{eqnarray*}
where $\alpha=2.68, r_0=0.258$ for $r\le r_1$ and  $\alpha=5.53,
r_0=-0.308$ for $r > r_1$ with
$$r_1=0.6(1+0.02\cos{(20\theta)})$$
for density and
$$r_1=0.62324965 $$ for pressure.
In this case,  two different isothermal equilibrium states intersect
around $r=0.6$ with density and pressure perturbations.

Due to the geometric symmetry in this problem,  only the solution in
the first quadrant is computed. The computational domain, $[0,1]
\times [0,1]$, is covered by $120 \times 120$ uniform grid points.
The CFL number takes a value $0.6$. On the axis of symmetry, a
rotated periodic boundary condition is used. At the other
boundaries, the values in the ghost cells are fixed as their initial
values in order to keep the isothermal equilibrium solution there.
Figure \ref{RT1} presents the time evolutions of the density
distributions at times $t=0.0, 0.8, 1.4,$ and $2.0$. Figure
\ref{RT2} shows a scatter plot of the density as a function of the
radius. Note that, away from the location ($r \simeq r_0$), the
current method retains the hydrostatic equilibrium solution and
radial symmetry very well.

\section{Conclusions}

The gas-kinetic BGK scheme  for the Navier-Stokes equations is
extended to include external gravitational forces. Our study allows
us to draw the following conclusions. (1) The gravitational force
effect should be included in the evaluation of the numerical fluxes
across a cell interface.  It has  the first-order effect in time
\cite{xu3}, which cannot be neglected in a second-order scheme. In a
gas kinetic scheme, the external force modifies the time evolution
of both the initial gas distribution function and the local
equilibrium state through the particle acceleration. Without such an
implementation, it is impossible for the system to settle down to a
highly accurate hydrostatic equilibrium state. (2) Numerical heating
is mainly caused by the inconsistent treatment of the gravitational
source terms inside each cell in the energy source term scheme.
After a long time evolution, the numerical error will accumulate and
dramatically affect the final solution of an isolated or insulated
gravitational system. For a system that exchanges energy with its
surrounding environment or that has a short evolution time, such as
the star explosion case, this kind of numerical heating may not
appear to be significant. (3) The numerical simulation of the onset
of two-dimensional laminar convection shows that the current BGK
scheme can accurately determine the critical Rayleigh numbers.
However, the nonlinear limiter is usually harmful to the smooth flow
solution. High-order interpolation may alleviate this problem. (4)
The results of finite amplitude laminar convection with a central
interpolation agree well with Graham's results. The van Leer limiter
introduces a bit of additional dissipation. By including the
tangential gradient terms, i.e., the so-called multidimensional
approach, the accuracy of the scheme is improved. The kinetic scheme
presented in this paper is not only limited to hydrodynamics under
gravitational fields; it can be applied to other systems in which
the external force term can be an electric field force, such as the
Boltzmann-type equations for semiconductor devices.

Developing a well-balanced scheme for gas dynamic equations under
gravitational fields is still an open problem. It is much more
difficult than developing a scheme for the shallow water equations.
We hope that this paper will help to shift attention from shallow
water equations to gas dynamic equations. At the current stage, it
seems that no currently available scheme can maintain the velocity
of the hydrostatic equilibrium state up to the machine zero, such as
on the order of $10^{-16}$. At the same time, the scheme has a shock
capturing ability. Even with consideration of all the techniques
presented in this paper, the current method has only second-order
accuracy in maintaining the hydrostatic solution. Much work is left
for the future.

\section*{Acknowledgments}
Part of our 2D and 3D computations were performed in the
Computational Clusters Laboratory at the Department of Mathematics
at the Hong Kong University of Science and Technology. The current
research was supported by  National Science Foundation of China
grants 10573022, 10337060, Hong Kong Research Grants Council grants
HKUST621005, 6119/02P, and  the Croucher Foundation.

\section*{Appendix: Solution of Matrix Equation $\mathbf{b}=\mathbf{Ma}$ in
the three-dimensional Case}

In the gas-kinetic scheme,  to obtain the derivatives of a
Maxwellian, the solution of the following equations need to be
evaluated,
 \begin{eqnarray} \label{apb1}\left(\begin{array}{ccc} b_1\\ b_2
\\b_3\\b_4\\b_5 \end{array}\right)=\mathbf{M}
\left(\begin{array}{ccc} a_1\\ a_2 \\a_3\\a_4\\a_5
\end{array}\right),
 \end{eqnarray} where $\mathbf{b}$ and $\mathbf{M}$ are
known. The matrix $\mathbf{M}$ is defined as
\begin{displaymath}
{\mathbf{M}}=\frac{1}{\rho}\int {{\vec{\psi}}\otimes {\vec{\psi}}}
f^{eq} d \Xi,
\end{displaymath}
where $f^{eq}$ is the Maxwellian.\\
The details of the matrix $\mathbf{M}$ are given by \\
\begin{displaymath}
\mathbf{M}= \left(\begin{array}{ccccc}
1 & u_1 & u_2 & u_3 & B_1\\
u_1 & u_1^2+1/2\lambda & u_1u_2 & u_1u_3 & B_2\\
u_2 & u_1u_2 & u_2^2+1/2\lambda & u_2u_3 & B_3\\
u_3 & u_1u_3 & u_2u_3 & u_3^2+1/2\lambda & B_4\\
B_1 & B_2 & B_3 & B_4 & B_5
\end{array}\right),
\end{displaymath}
where
\begin{eqnarray}
B_1&=&\frac{1}{2}(u_1^2+u_2^2+u_3^2+(K+3)/2\lambda),\nonumber\\
B_2&=&\frac{1}{2}(u_1^3+u_1(u_2^2+u_3^2)+(K+5)u_1/2\lambda),\nonumber\\
B_3&=&\frac{1}{2}(u_2^3+u_2(u_1^2+u_3^2)+(K+5)u_2/2\lambda),\nonumber\\
B_4&=&\frac{1}{2}(u_3^3+u_3(u_1^2+u_2^2)+(K+5)u_3/2\lambda),\nonumber
\end{eqnarray}
and
\begin{eqnarray}
B_5=\frac{1}{4}((u_1^2+u_2^2+u_3^2)^2+(K+5)(u_1^2+u_2^2+u_3^2)/\lambda
+(K+3)(K+5)/4\lambda^2).\nonumber
\end{eqnarray}
Defining
\begin{eqnarray}
R_5&=&2b_5-\left(u_1^2+u_2^2+u_3^2+\frac{K+3}{2\lambda}\right)b_1,\nonumber\\
R_4&=&b_4-u_3 b_1,\nonumber\\
R_3&=&b_3-u_2 b_1,\nonumber\\
R_2&=&b_2-u_1 b_1,\nonumber
\end{eqnarray}
the solutions of Eq.(\ref{apb1}) are
\begin{eqnarray}
a_5&=&\frac{4\lambda^2}{K+3}(R_5-2u_1R_2-2u_2R_3-2u_3R_4),\nonumber\\
a_4&=&2\lambda R_4-u_3a_5,\nonumber\\
a_3&=&2\lambda R_3-u_2a_5,\nonumber\\
a_2&=&2\lambda R_2-u_1a_5,\nonumber
\end{eqnarray}
and
\begin{displaymath}
a_1=b_1-u_1a_2-u_2a_3-u_3a_4-\frac{1}{2}a_5\left(u_1^2+u_2^2+u_3^2+\frac{K+3}{2\lambda}\right).
\end{displaymath}

\clearpage

\begin{table}
      \caption[]{Critical Rayleigh number.}
         \label{crn}
     $$
     \begin{tabular}{c c c c c c}
            \hline\hline
            \noalign{\smallskip}
            Z    &  0.5 & 1 &2 &4 &10\\
            \noalign{\smallskip}
            \hline
            \noalign{\smallskip}
            $Ra^{\mathrm{a}}$ & 545.03 &422.79 &288.51&164.88 &58.17\\
            $Ra^{\mathrm{b}}$ & 550.02 &424.84 &289.82&167.24 &59.96\\
            $\delta^{\mathrm{c}}$ & 0.4 &0.2 &0.1&0.07 &0.02\\
            \noalign{\smallskip}
            \hline
\end{tabular}
     $$
\begin{list}{}{}
\item[$^{\mathrm{a}}$] Linear analysis results from Graham
\cite{graham}. \item[$^{\mathrm{b}}$] Results from current
gas-kinetic BGK scheme. \item[$^{\mathrm{c}}$]Courant number.
\end{list}
   \end{table}

\begin{table}
      \caption[]{Maximum Mach number ($Mc$) vs. relative Rayleigh Number ($R^{\ast}$).}
         \label{mc-rn}
     $$
     \begin{tabular}{c c c c c c c}
            \hline\hline
            \noalign{\smallskip}
            $R^{\ast}$ &  3 & 7 &10&  30& 70 &100\\
            \noalign{\smallskip}
            \hline
            \noalign{\smallskip}
            $Mc^{\mathrm{a}}$ & 0.072 &0.097 &0.107& 0.136 & 0.161& 0.172\\
            $Mc^{\mathrm{b}}$ & 0.071 &0.096 &0.106& 0.134 & 0.156& 0.166\\
            $Mc^{\mathrm{c}}$ & 0.067 &0.095 &0.104& 0.131 & 0.151& $0.159^\ast$\\
            \noalign{\smallskip}
            \hline
\end{tabular}
     $$
\begin{list}{}{}
\item[$^{\mathrm{a}}$] Current BGK scheme with central interpolation.
\item[$^{\mathrm{b}}$] Current BGK scheme with the van Leer limiter interpolation.
\item[$^{\mathrm{c}}$] Dimensional-splitting BGK scheme with the van
Leer limiter. \item[$^{\ast}$] Temporally averaged value.
\end{list}
\end{table}

\begin{table}
      \caption[]{ Nusselt number ($\cal N$) vs. relative Rayleigh Number ($R^{\ast}$).}
         \label{nu-rn}
     $$
     \begin{tabular}{c c c c c c c}
            \hline\hline
            \noalign{\smallskip}
            $R^{\ast}$ &  3 & 7 &10&  30& 70 &100\\
            \noalign{\smallskip}
            \hline
            \noalign{\smallskip}
            ${\cal N}^{\mathrm{a}}$ & 2.575 &3.729 &4.334& 6.938 & 10.000& 11.676\\
            ${\cal N}^{\mathrm{b}}$ & 2.533 &3.702 &4.299& 6.831 & 9.675& 11.141\\
            ${\cal N}^{\mathrm{c}}$ & 2.236 &3.393 &3.989& 6.777 & $9.849^\ast$& $11.252^\ast$\\
            \noalign{\smallskip}
            \hline
\end{tabular}
     $$
\begin{list}{}{}
\item[$^{\mathrm{a}}$] Current BGK scheme with central interpolation.
\item[$^{\mathrm{b}}$] Current BGK scheme with the van Leer limiter.
\item[$^{\mathrm{c}}$] Dimensional-splitting BGK scheme with the van
Leer limiter. \item[$^{\ast}$] Temporally averaged values.
\end{list}
\end{table}

\clearpage

\begin{figure}
     \centerline{\psfig{figure=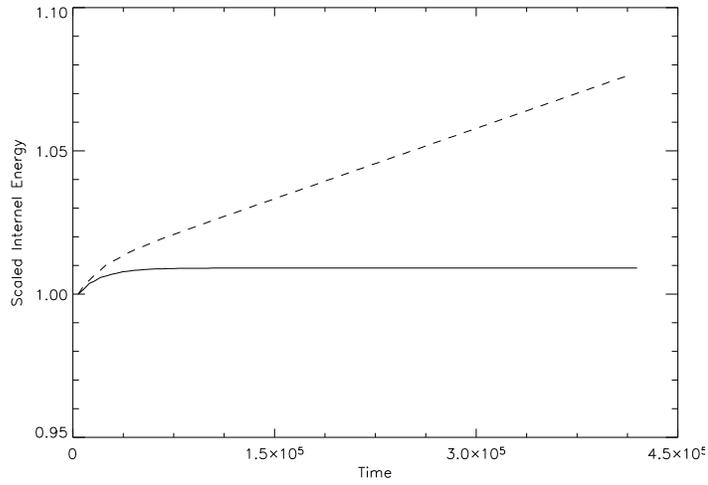,height=2.7in,angle=0,clip=}}
      \caption{Numerical heating. The time evolution of
      the total internal energy of an isolated  system.
      Both solutions are computed by the BGK scheme but with different source term
      treatment. The total internal energy is scaled by the initial
      value. The dashed line is the solution from the BGK scheme without including the
      source term effect in the flux function and the solid line is the solution
      with the source term effect included in the flux.
     }
\label{Figheat}
   \end{figure}

\begin{figure}
   \centering
   \centerline{\psfig{figure=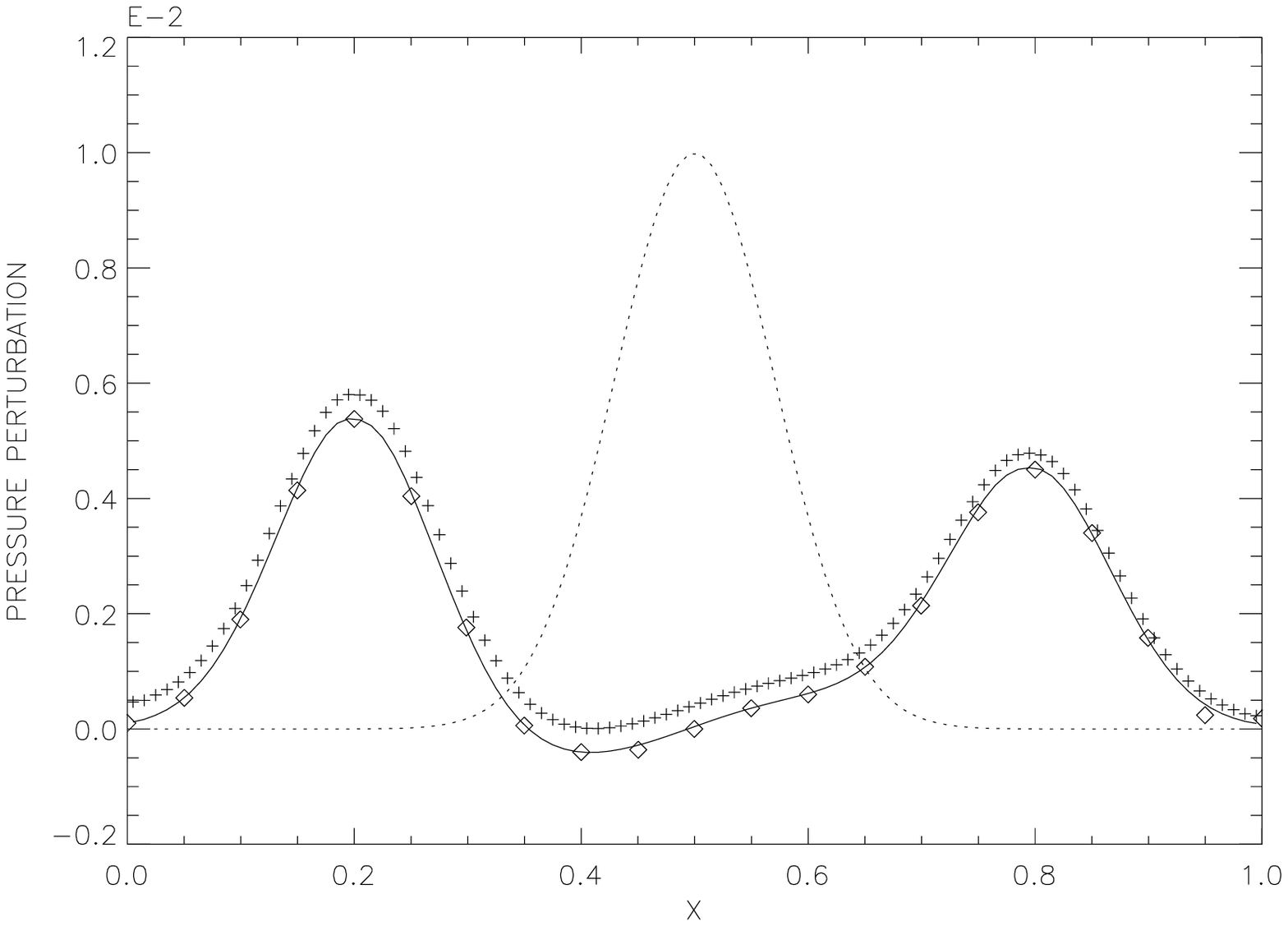,height=2.7in,angle=0,clip=}}
   \centerline{\psfig{figure=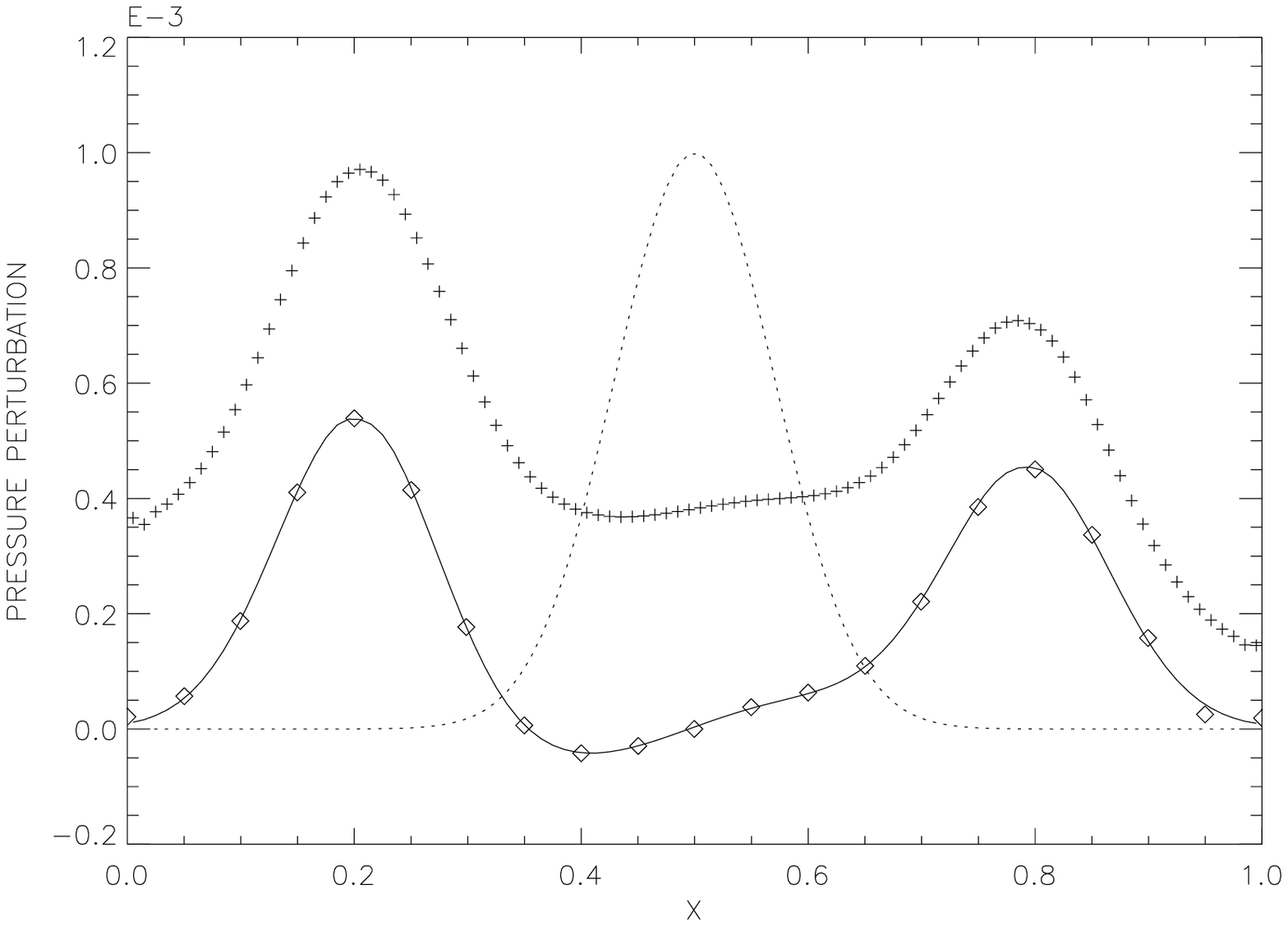,height=2.7in,angle=0,clip=}}
      \caption{Solutions of the BGK scheme with and without the source term effect in the
      flux function. The BGK scheme with the van Leer limiter is used in
      these
      calculations. $100$ grid points are used in the whole domain. The
      output time is $t=0.25$.
      The upper and lower figures have the perturbations; upper: $\eta=0.01$, lower: $\eta =0.001$.
       The dotted lines show the initial pressure perturbation from a hydrostatic solution,
       the pluses
      represent the results without the source term effect in the flux, and the solid lines are
      the results with the source term effect. The diamonds are the reference solutions from
      \cite{leveque}.}
         \label{FigLe1}
   \end{figure}

\begin{figure}
   \centering
   \centerline{\psfig{figure=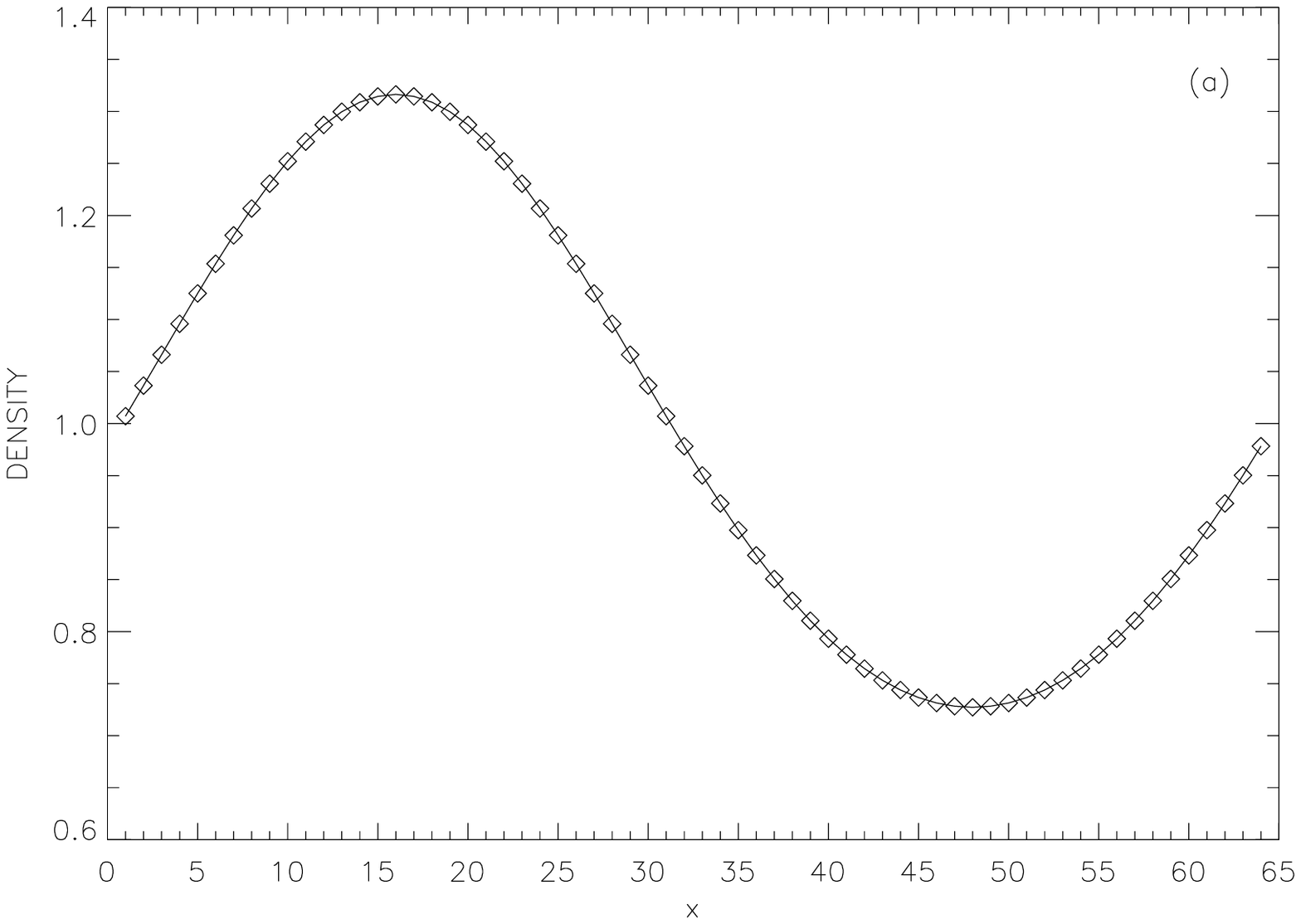,height=2.7in,angle=0,clip=}}
   \centerline{\psfig{figure=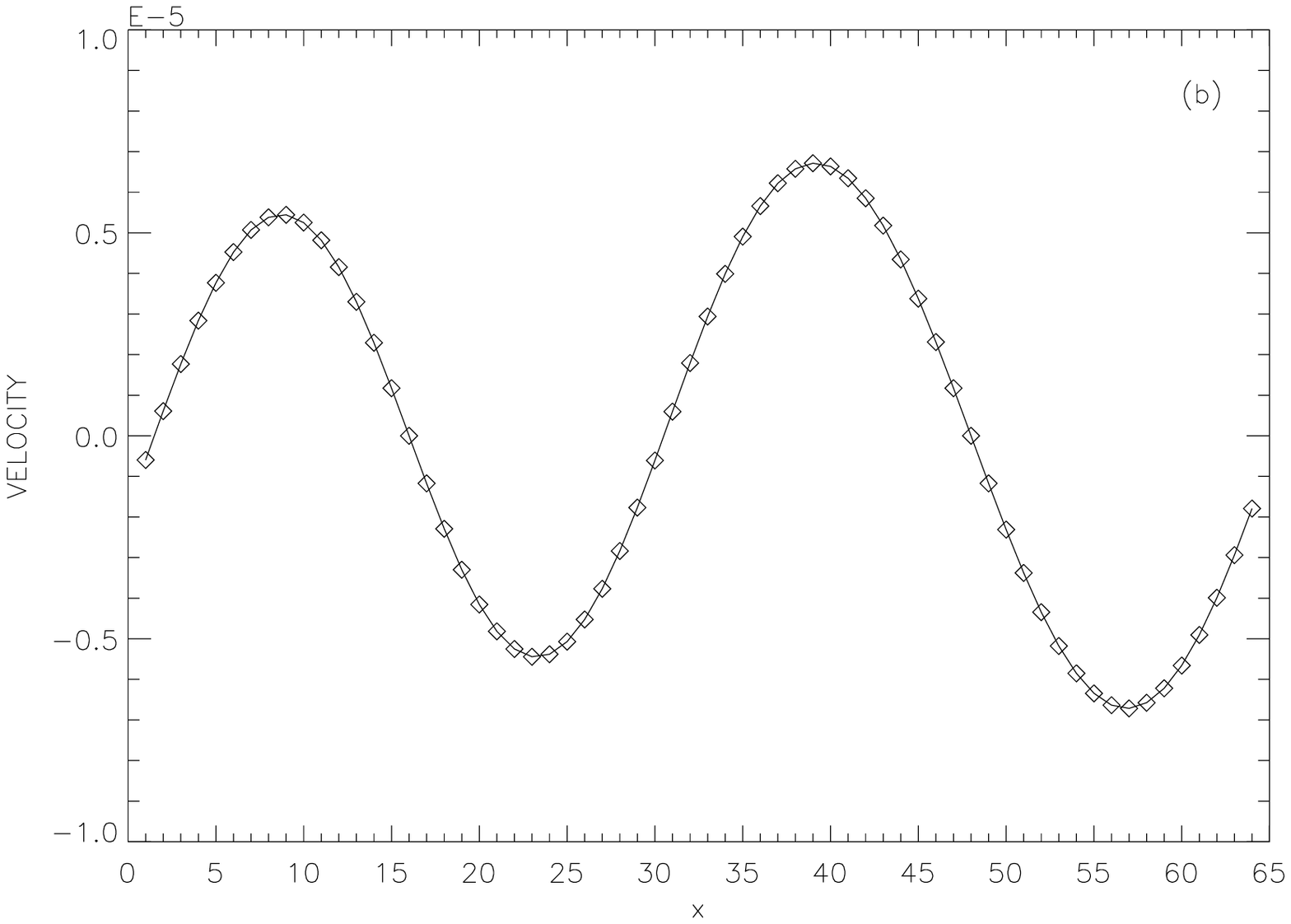,height=2.7in,angle=0,clip=}}
   \centerline{\psfig{figure=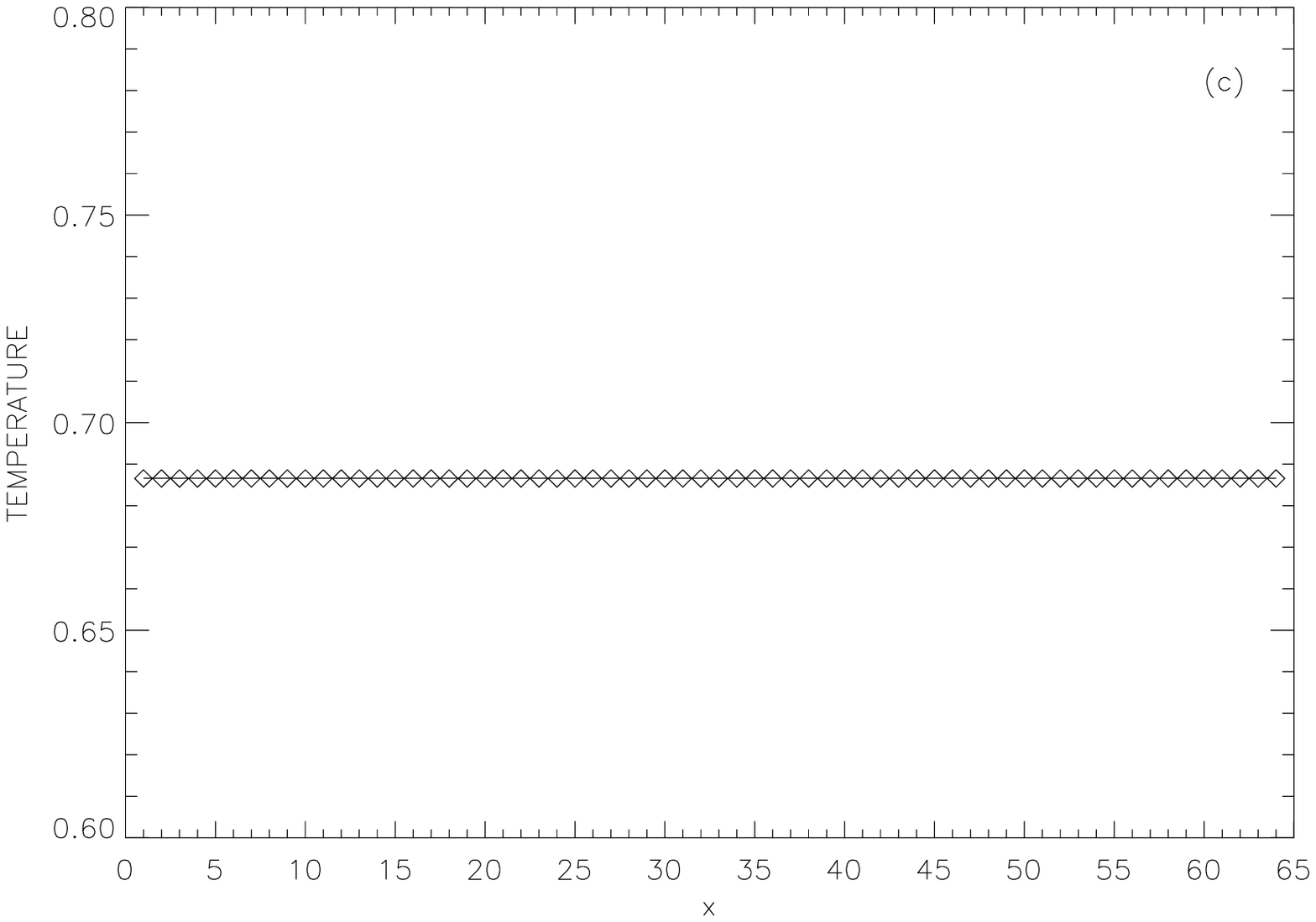,height=2.7in,angle=0,clip=}}
      \caption{Comparison of the EST (Diamonds) and ECT (Solid lines) schemes.
       These results were obtained after $500,000$ time steps.}
         \label{FigUe1}
   \end{figure}

\begin{figure}
    \centerline{\psfig{figure=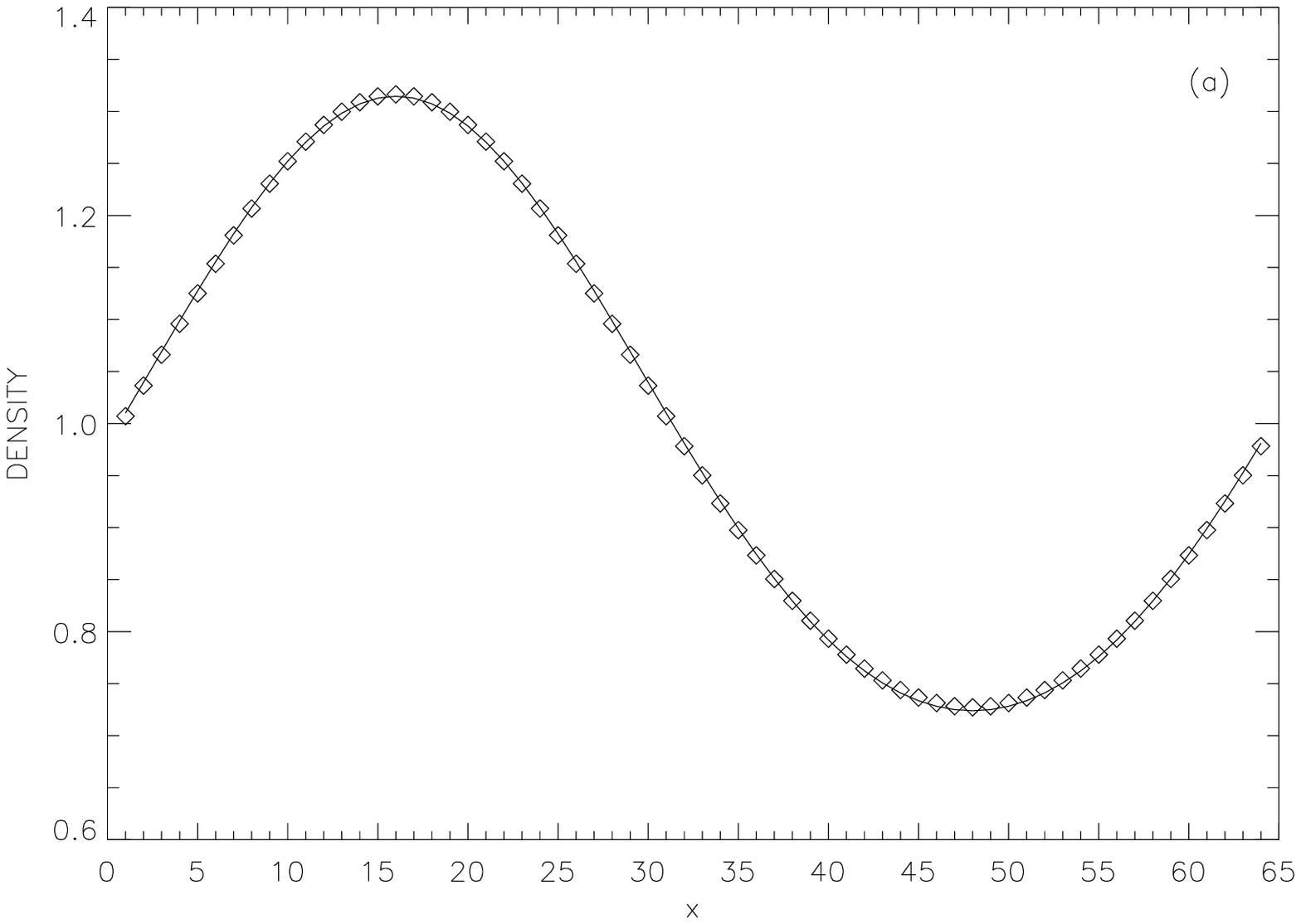,height=2.7in,angle=0,clip=}}
    \centerline{\psfig{figure=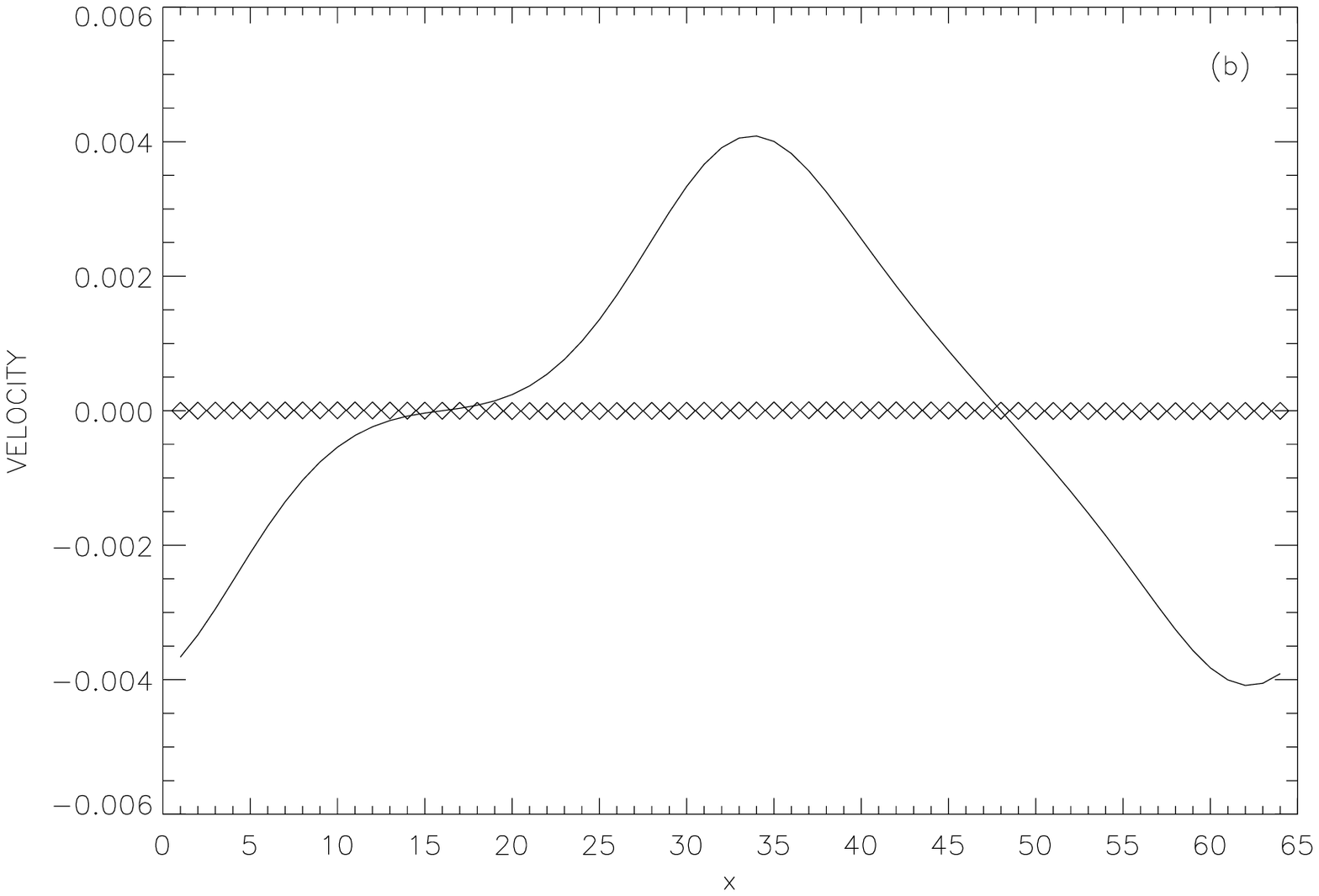,height=2.7in,angle=0,clip=}}
    \centerline{\psfig{figure=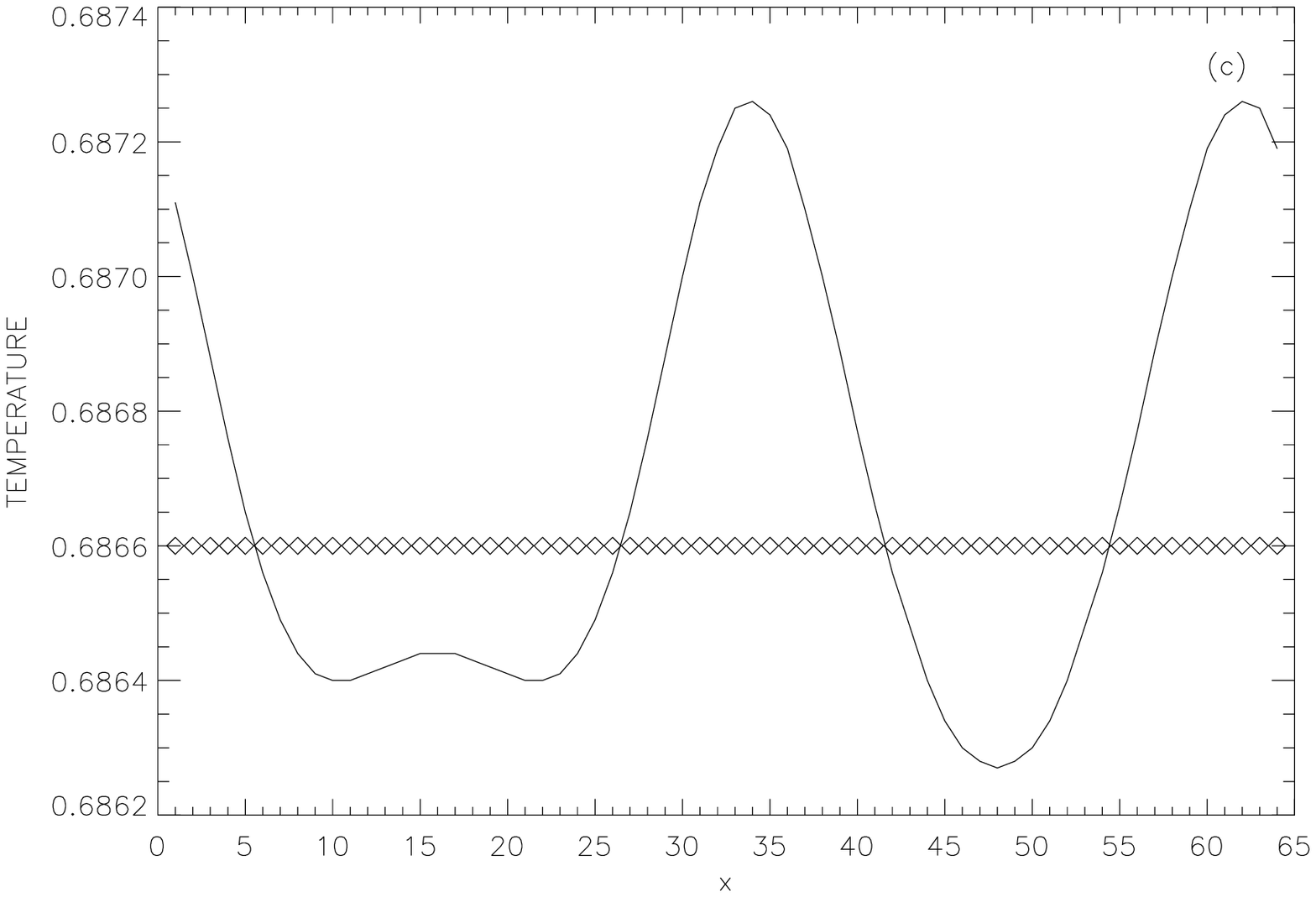,height=2.7in,angle=0,clip=}}
      \caption{Results of the ECT scheme with the external force term in the flux (diamonds), and
      the ECT scheme without including external force term in the flux (solid lines).
       These results were obtained after $500,000$ time steps.}
         \label{FigUe2}
   \end{figure}

\begin{figure}
 \centering
   \centerline{\psfig{figure=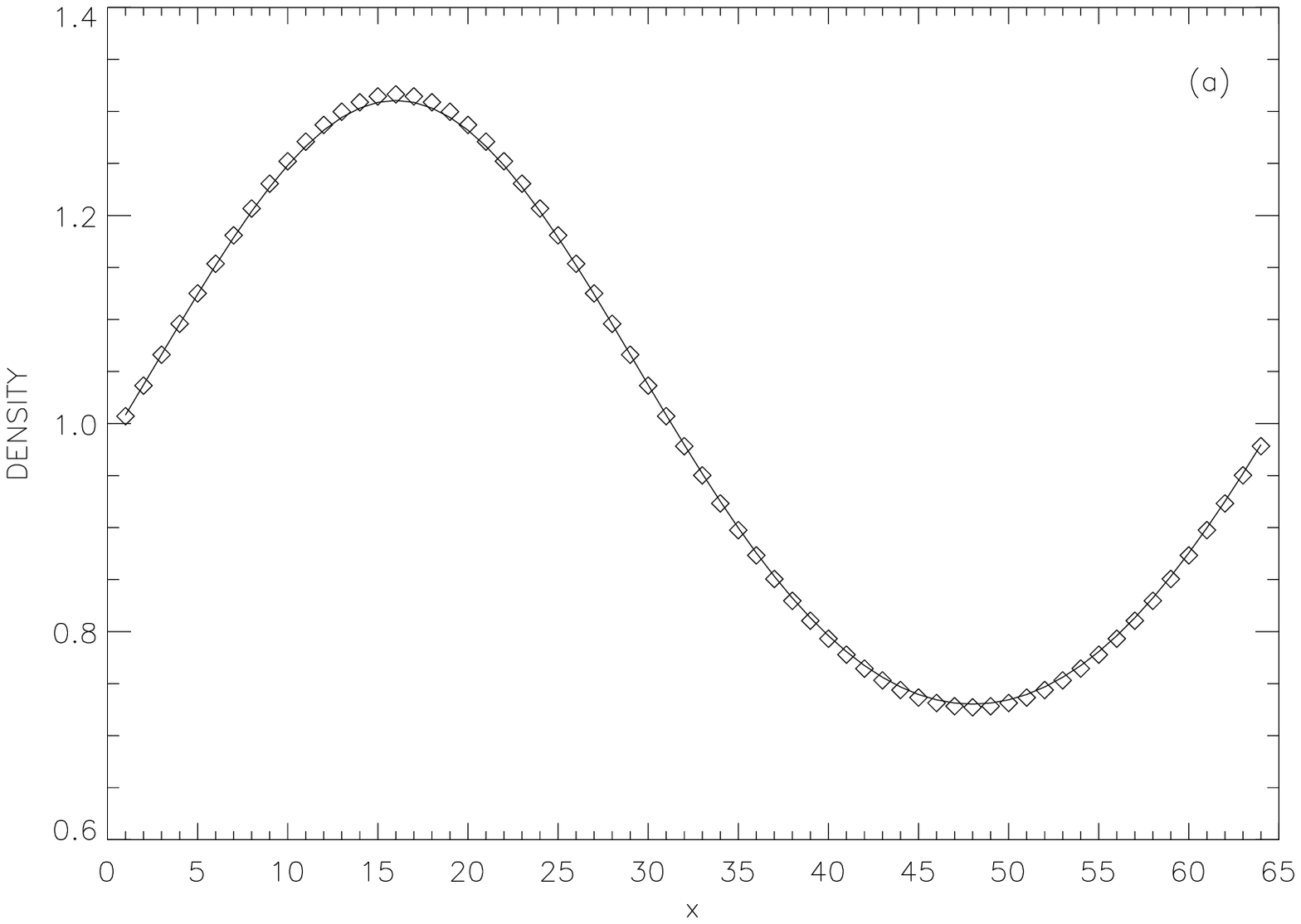,height=2.7in,angle=0,clip=}}
    \centerline{\psfig{figure=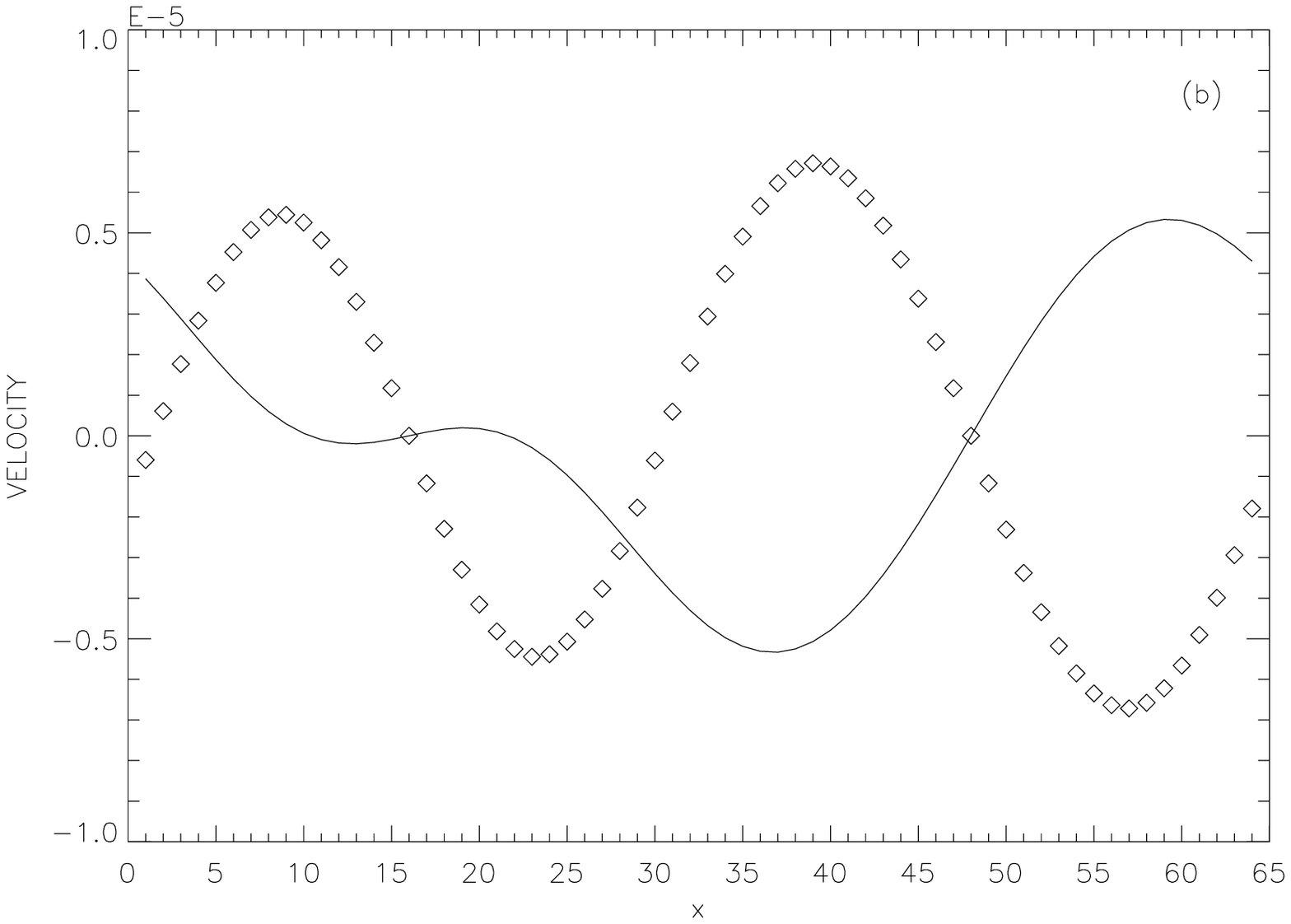,height=2.7in,angle=0,clip=}}
    \centerline{\psfig{figure=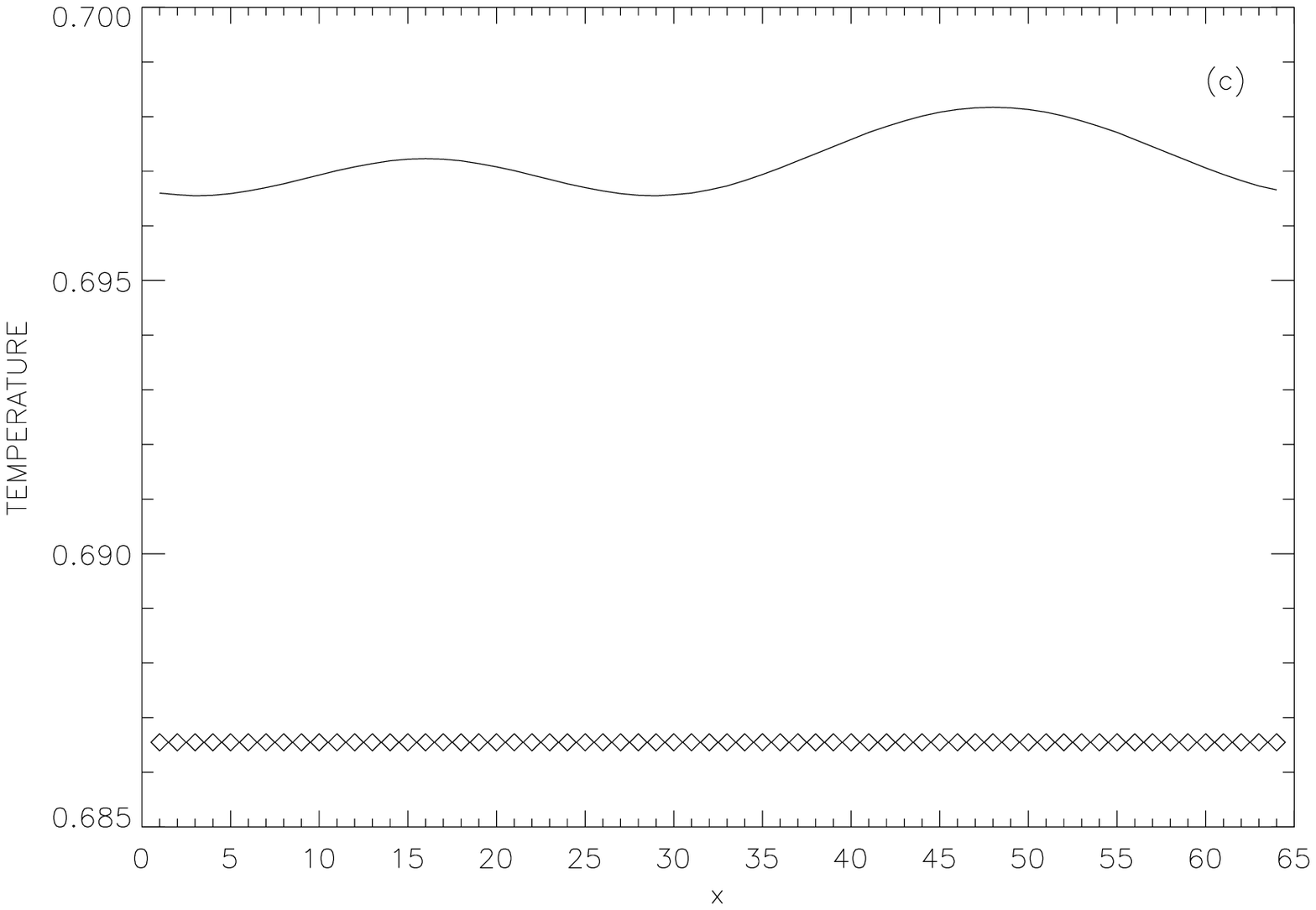,height=2.7in,angle=0,clip=}}
\caption{Results of the current EST BGK scheme (diamonds) and the
Strang splitting method (solid lines). The
 results were obtained after $500,000$ time steps.}
 \label{FigStrang2}
\end{figure}

\begin{figure}
   \centerline{\psfig{figure=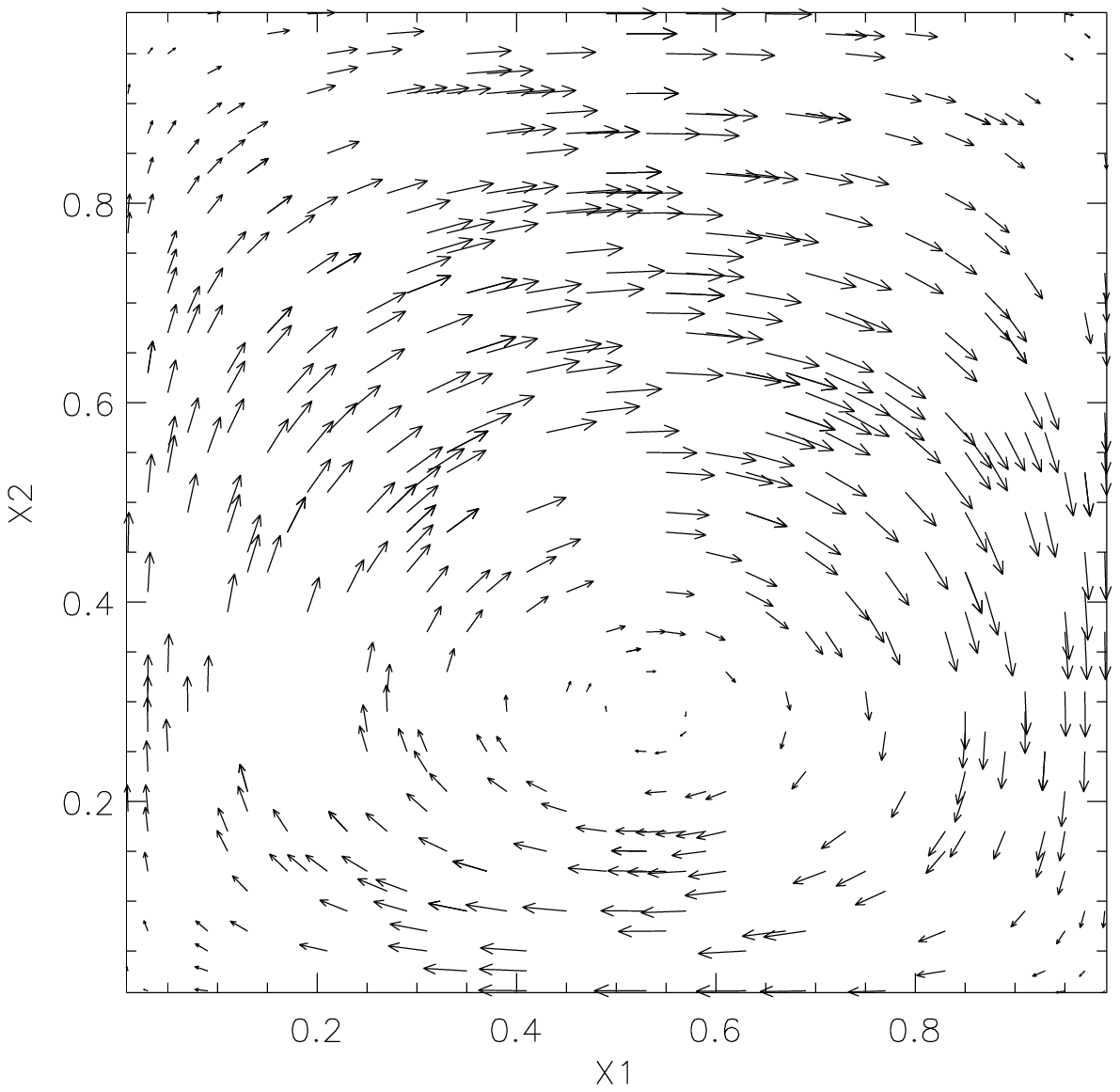,height=2.7in,angle=0,clip=}
   \psfig{figure=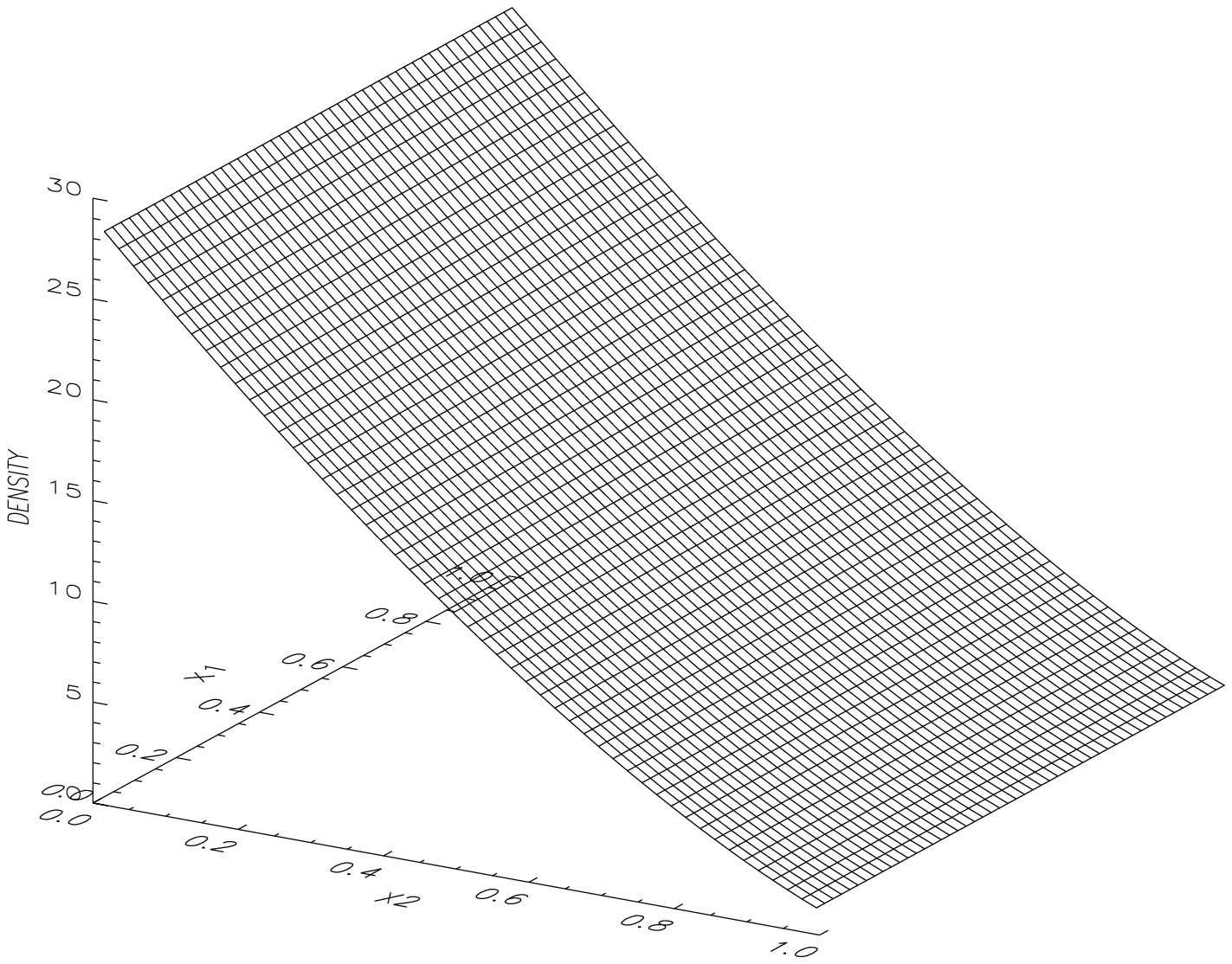,height=2.7in,angle=0,clip=}}
   \centerline{\psfig{figure=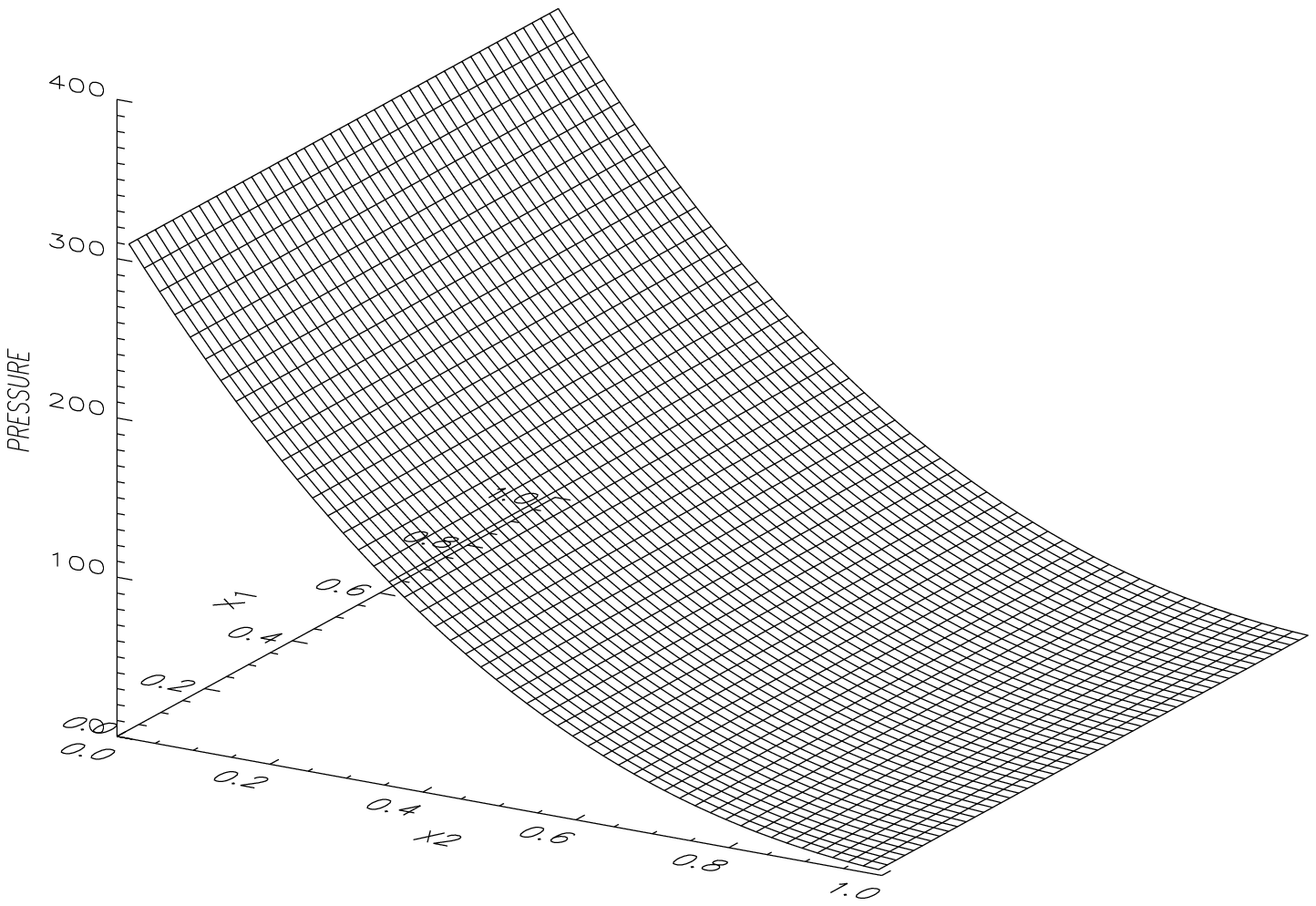,height=2.7in,angle=0,clip=}
   \psfig{figure=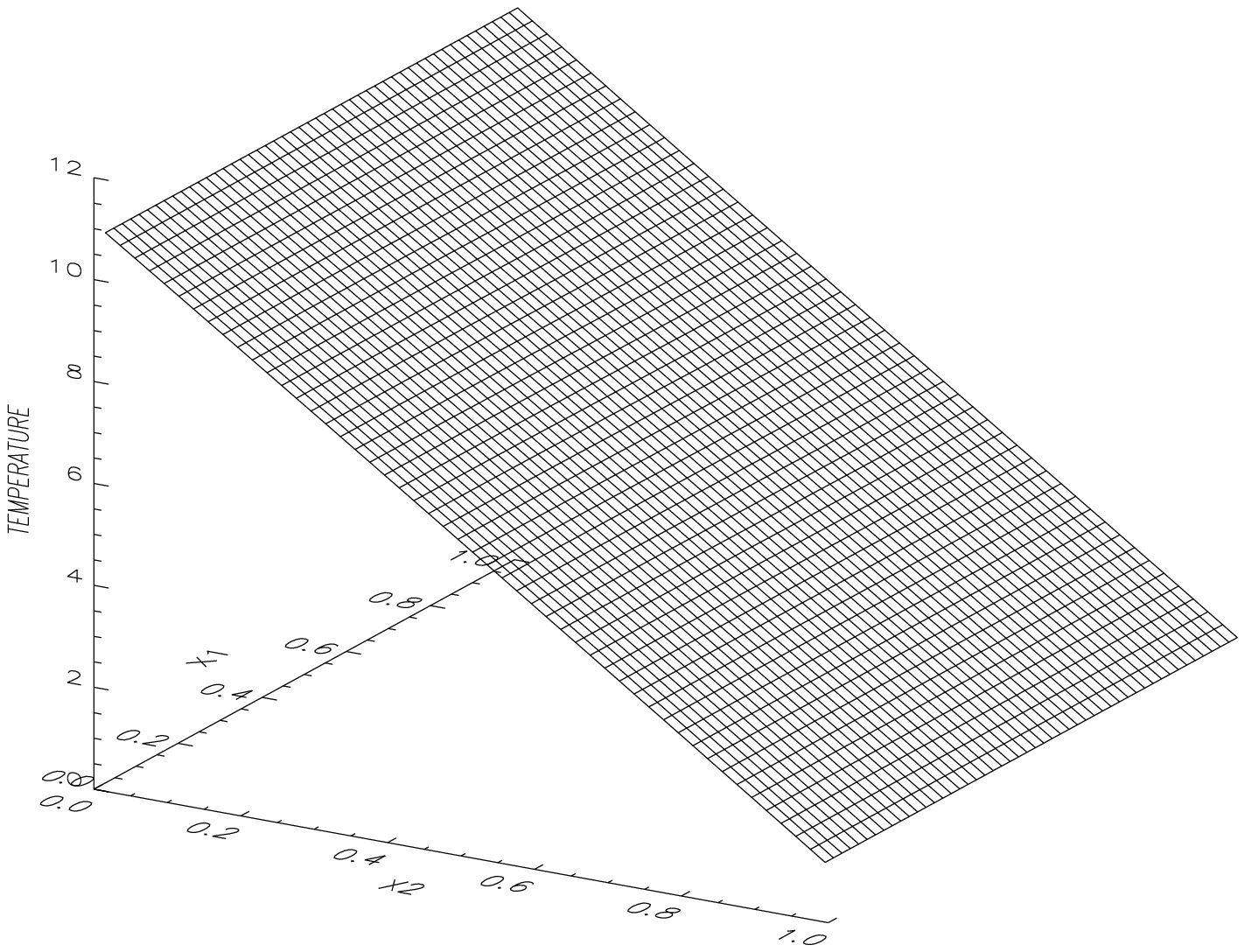,height=2.7in,angle=0,clip=}}
 \caption{Solutions of two-dimensional laminar convection obtained by the BGK scheme.
      The upper-left, upper-right, lower-left and lower-right plots show the velocity vector field, the
      density distribution, the pressure distribution, and
      temperature distribution, respectively. The corresponding
      parameters are
      $Z=10$ and $Ra=61.$}
 \label{Figsl}
\end{figure}

\begin{figure}
    \centerline{\psfig{figure=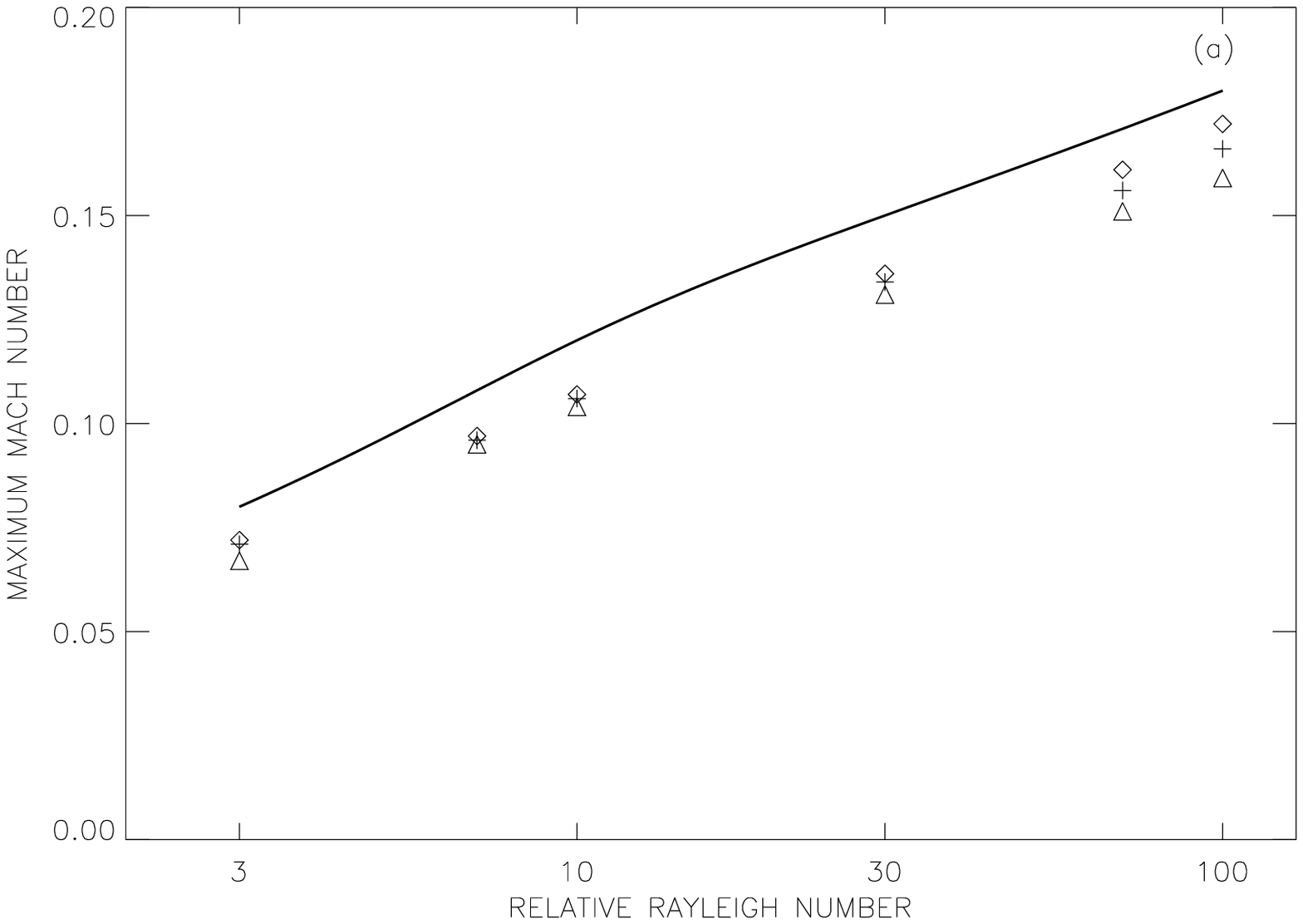,height=2.7in,angle=0,clip=}}
\centerline{\psfig{figure=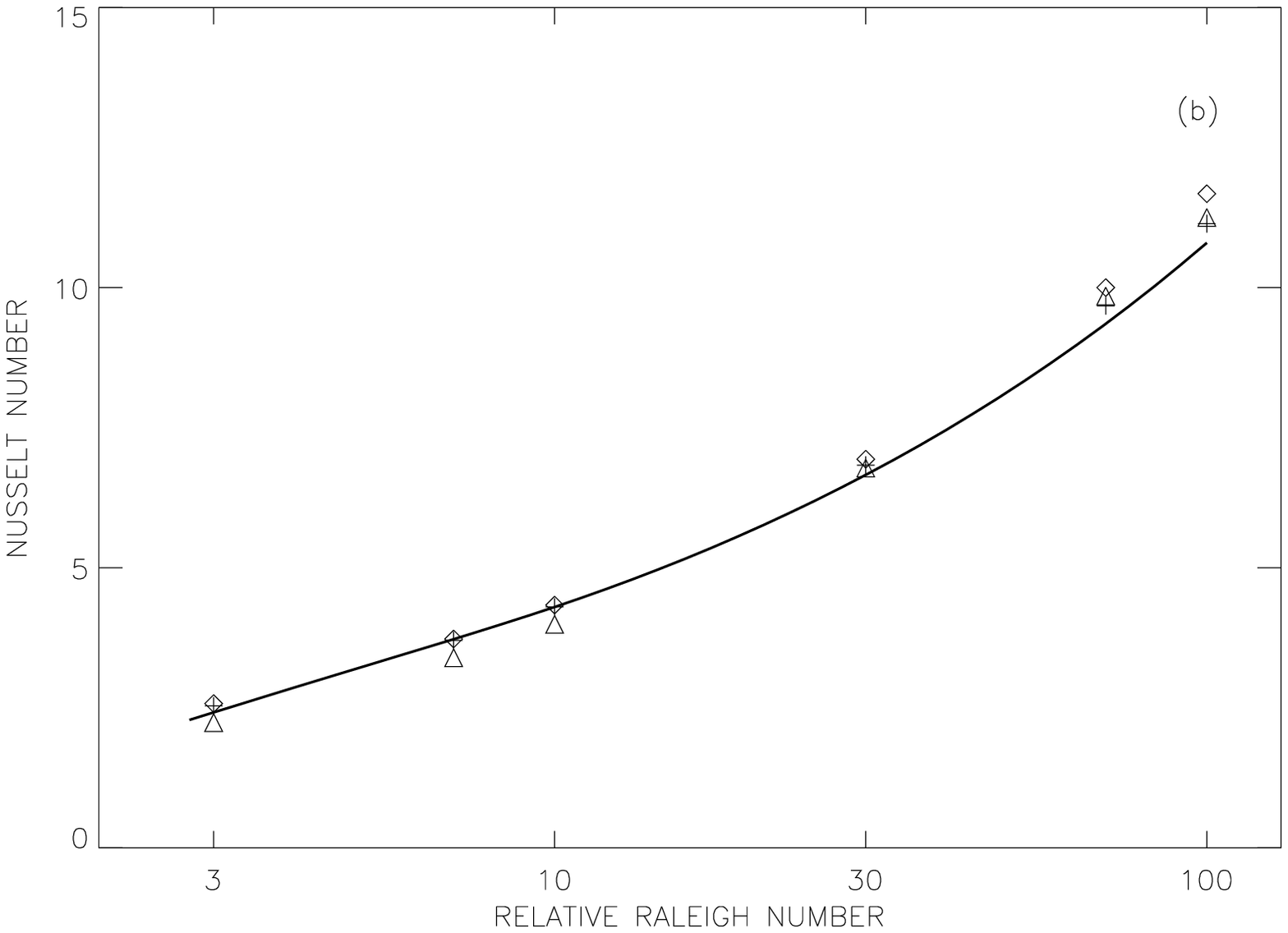,height=2.7in,angle=0,clip=}}
      \caption{Maximum Mach number (a) and Nusselt number (b)
      vs. relative Rayleigh numbers.
      The solid lines are the results
      from Graham's paper \cite{graham} and the diamonds and pluses
      represent the results computed by the BGK schemes
      with central interpolation and the van Leer limiter reconstruction.
      The triangles show the results
      from the dimensional-splitting BGK scheme with the van Leer limiter.}
         \label{Figramn}
   \end{figure}

\begin{figure}
\centerline{\psfig{figure=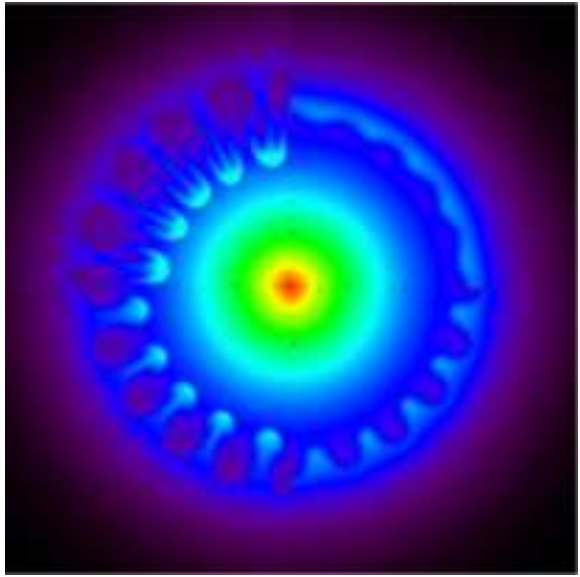,height=3.0in,angle=0,clip=}}
      \caption{A Rayleigh-Taylor instability with gravitational field directed radially inward.
      Density contours at four different times ($t=0.0, 0.8, 1.4$, and
      $2.0$) are shown in the four quadrants, starting with the initial data in the upper
      right corner and progressing clockwise.}
         \label{RT1}
   \end{figure}

\begin{figure}
\hskip -0.5cm  
\centerline{\psfig{figure=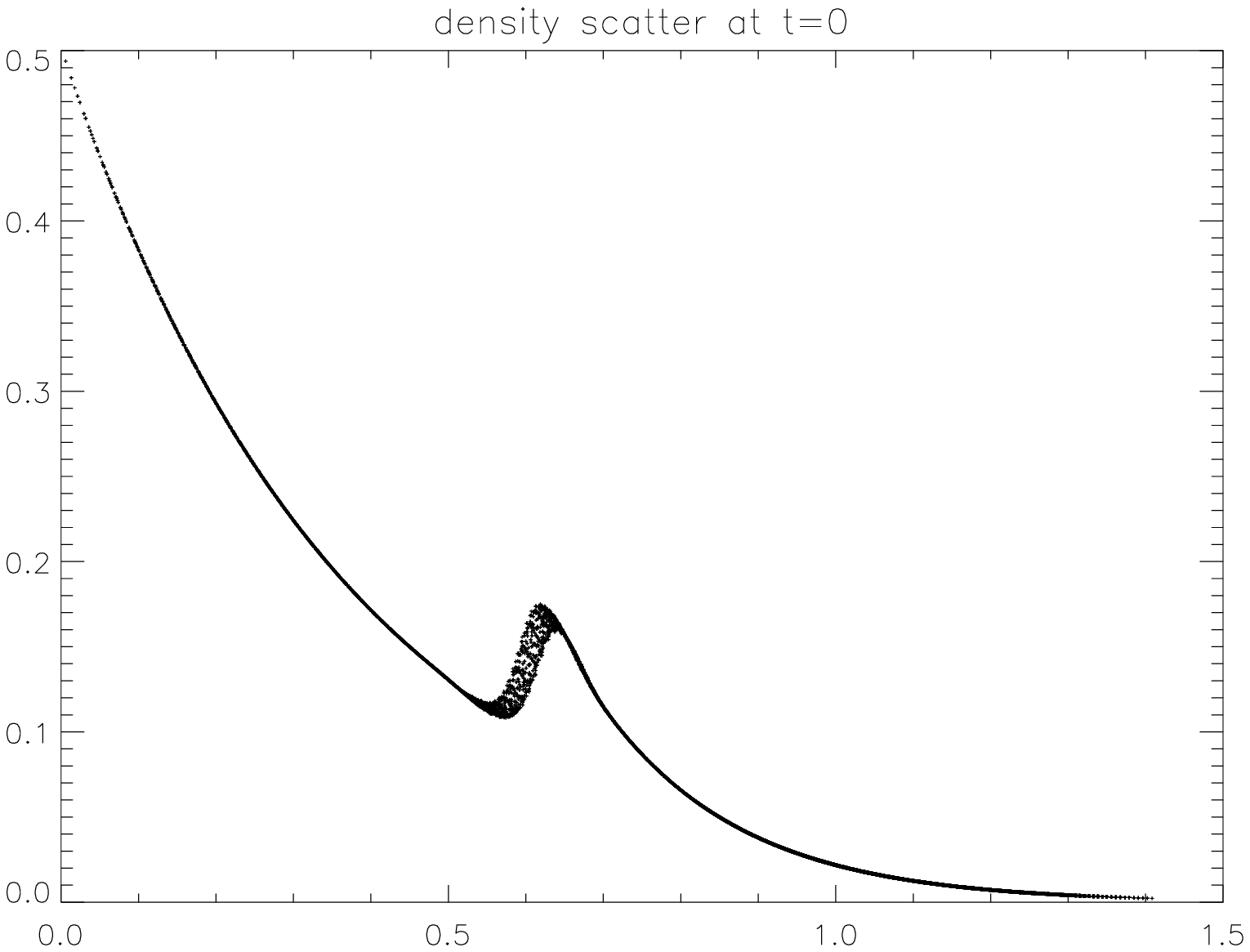,height=2.04in,angle=0,clip=}
    \psfig{figure=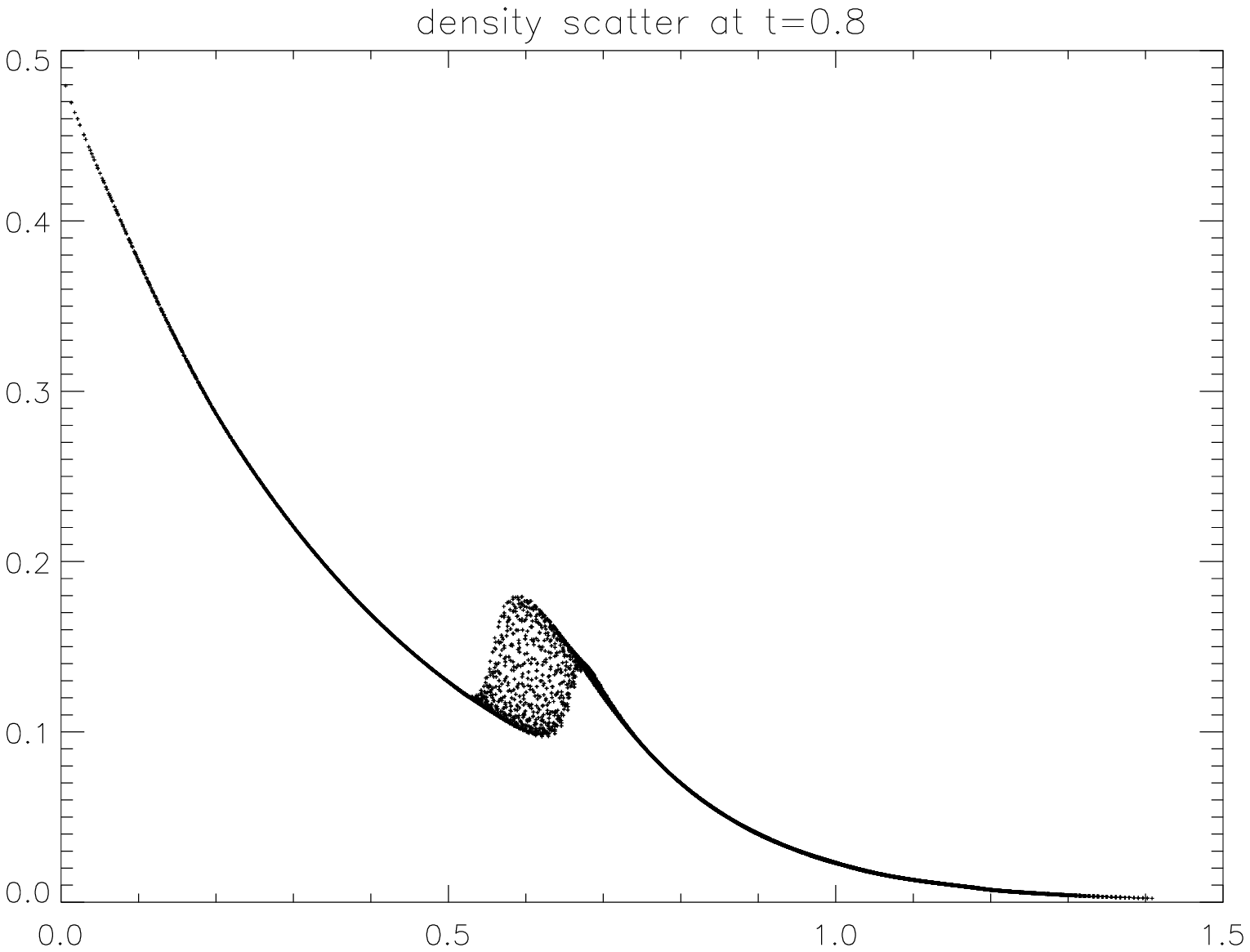,height=2.04in,angle=0,clip=}}
\centerline{\psfig{figure=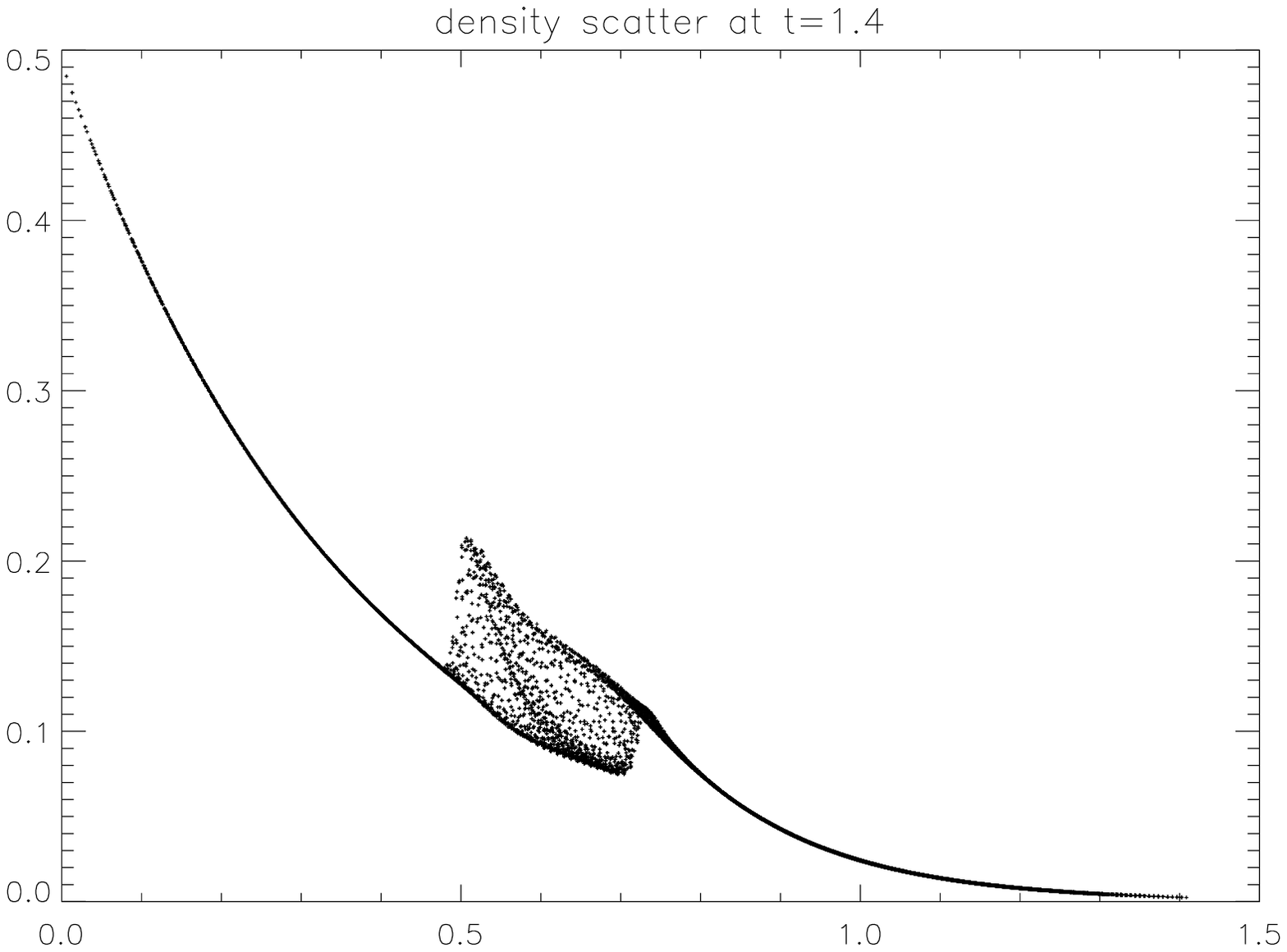,height=2.0in,angle=0,clip=}
    \psfig{figure=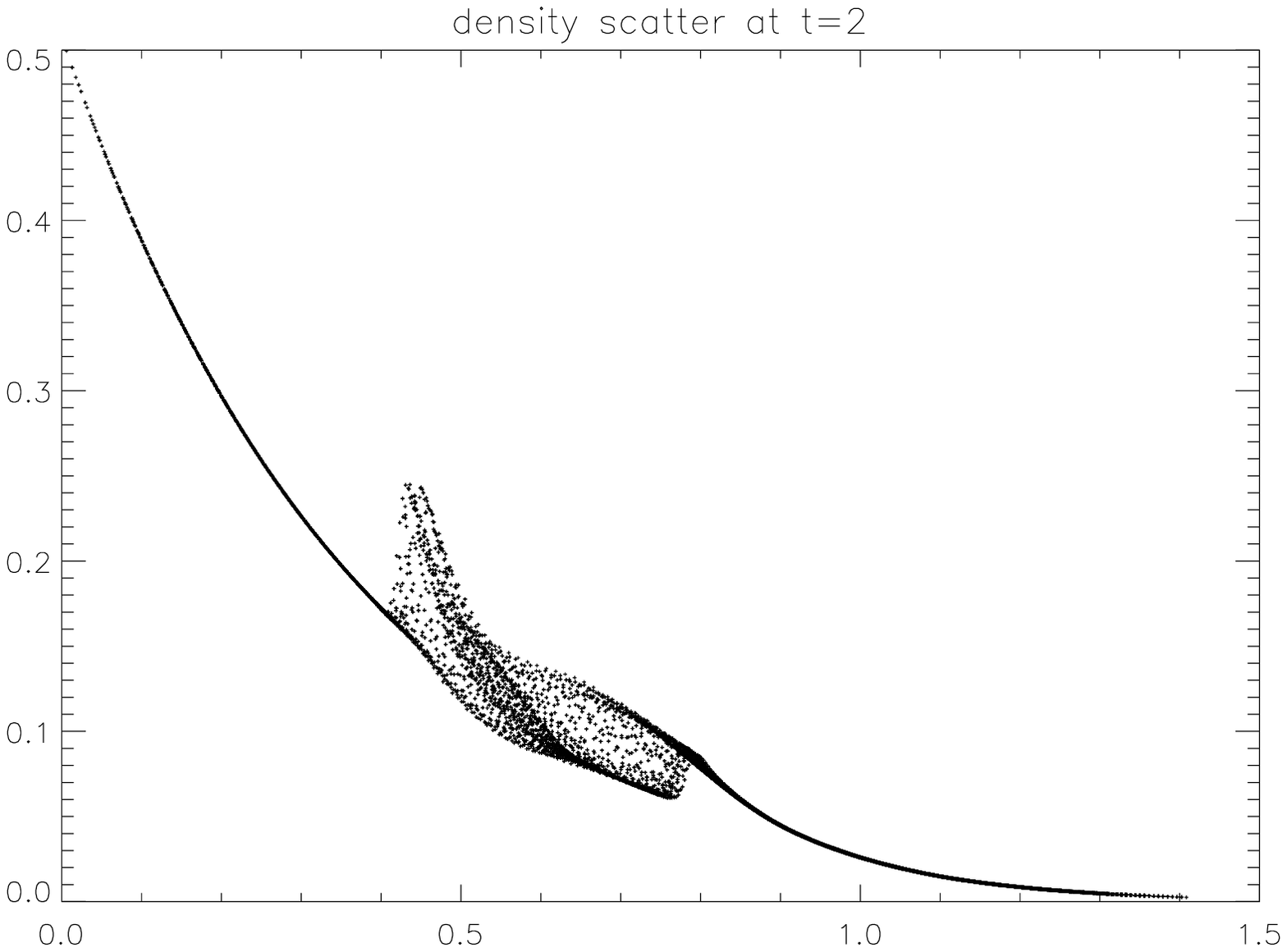,height=2.0in,angle=0,clip=}}
      \caption{Scatter plots of the density in the cell vs. the distance of the cell center from the origin.
      The density is shown at four different times ($t=0.0, 0.8, 1.4$, and
      $2.0$). The single line away from $r_0 $ indicates the capacity of the current scheme to
      keep the hydrostatic equilibrium solution.}
         \label{RT2}
   \end{figure}

\end{document}